\documentclass[sigconf, screen]{acmart}
\AtBeginDocument{%
  }

\copyrightyear{2026}
\acmYear{2026}
\setcopyright{cc}
\setcctype{by-nc-nd}
\acmConference[CHI '26]{Proceedings of the 2026 CHI Conference on Human Factors in Computing Systems}{April 13--17, 2026}{Barcelona, Spain}
\acmBooktitle{Proceedings of the 2026 CHI Conference on Human Factors in Computing Systems (CHI '26), April 13--17, 2026, Barcelona, Spain}
\acmDOI{10.1145/3772318.3790300}
\acmISBN{979-8-4007-2278-3/2026/04}
\newcommand{\revise}[1]{{#1}}
\newcommand{\rerevise}[1]{\textcolor{black}{#1}}

\AtBeginMaketitle{%
\let\oldAtBeginDocument\AtBeginDocument%
\renewcommand\AtBeginDocument[1]{#1}
\setcopyright{cc}
\setcctype{by-nc-nd}
\copyrightyear{2026}
\acmYear{2026}
\acmDOI{10.1145/3772318.3790300}
\acmISBN{979-8-4007-2278-3/26/04}
\acmConference[CHI '26]{Proceedings of the 2026 CHI Conference on Human Factors in Computing Systems}{April 13--17, 2026}{Barcelona, Spain}
\acmBooktitle{Proceedings of the 2026 CHI Conference on Human Factors in Computing Systems (CHI '26), April 13--17, 2026, Barcelona, Spain}
\let\AtBeginDocument\oldAtBeginDocument%
}

\begin{document}
\title{How Do Human Creators Embrace Human-AI Co-Creation? A Perspective on Human Agency of Screenwriters}

\author{Yuying Tang}
\orcid{0009-0003-1906-2834}
\affiliation{\institution{The Hong Kong University of Science and Technology}
\city{Hong Kong SAR}
\country{China}}
\email{yuying.tang@connect.ust.hk}

\author{Jiayi Zhou}
\orcid{0000-0003-4669-4872}
\affiliation{\institution{The Hong Kong University of Science and Technology}
\city{Hong Kong SAR}
\country{China}}
\email{jzhoudp@connect.ust.hk}

\author{Haotian Li}
\authornote{Haotian Li is the corresponding author.}
\orcid{0000-0001-9547-3449}
\affiliation{\institution{Microsoft Research Asia}
\city{Beijing}
\country{China}}
\email{haotian.li@microsoft.com}

\author{Xing Xie}
\orcid{0009-0009-3257-3077}
\affiliation{\institution{Microsoft Research Asia}
\city{Beijing}
\country{China}}
\email{xingx@microsoft.com}

\author{Xiaojuan Ma}
\orcid{0000-0002-9847-7784}
\affiliation{\institution{The Hong Kong University of Science and Technology}
\city{Hong Kong SAR}
\country{China}}
\email{mxj@cse.ust.hk}

\author{Huamin Qu}
\orcid{0000-0002-3344-9694}
\affiliation{\institution{The Hong Kong University of Science and Technology}
\city{Hong Kong SAR}
\country{China}}
\email{huamin@cse.ust.hk}

\begin{abstract}

Generative AI has greatly transformed creative work in various domains, such as screenwriting. To understand this transformation, prior research often focused on capturing a snapshot of human-AI co-creation practice at a specific moment, with less attention to how humans mobilize, regulate, and reflect to form the practice gradually. Motivated by Bandura's theory of human agency, we conducted a two-week study with 19 professional screenwriters to investigate how they embraced AI in their creation process. Our findings revealed that screenwriters not only mindfully planned, foresaw, and responded to AI usage, but, more importantly, through reflections on practice, they developed themselves and human-AI co-creation paradigms, such as cognition, strategies, and workflows. They also expressed various expectations for how future AI should better support their agency. Based on our findings, we conclude this paper with extensive discussion and actionable suggestions to screenwriters, tool developers, and researchers for sustainable human-AI co-creation.

\end{abstract}

\maketitle

%%%% 01-introduction.tex starts here %%%%

\section{Introduction}\label{sec:intro}

Since the public release of ChatGPT in November 2022 \cite{NewYorkTimes}, generative AI models such as large language models (LLMs) and diffusion models have rapidly reshaped creative industries~\cite{xie2025social}. Screenwriting, in particular, has become a frontline site of both experimentation and contestation. On the one hand, screenwriters have eagerly tested AI's potential to accelerate ideation and drafting~\cite{tang2025understanding}; on the other hand, anxieties over authorship, labor, and creative integrity escalated into collective action, most notably the 2023 Writers Guild of America strike~\cite{nyt_wga_strike_ai, 10.1145/3716135}. Now, three years later, the question is no longer whether AI will enter creative practice, but rather \textit{how creative workers are adapting to and developing their co-creation with AI}. \revise{The increasing integration of AI in screenwriting transforms not only the quality and style of final outputs~\cite{10.1145/3172944.3172972, 10.1145/3544548.3581225} but also the practices through which story ideas are conceived and developed~\cite{sanghrajka2017lisa, kapadia2015computer, tapaswi2014storygraphs, kim2017visualizing}. Building on this shift, prior studies have examined how writers interact with AI during co-creation, describing their active strategies, roles, and workflow patterns~\cite{tang2025understanding, 10.1145/3544548.3580782}. However, these accounts focus on isolated stages of use and therefore provide only snapshots of human–AI collaboration. 
% For example, a recent study by Tang et al. shows 
% that AI can stimulate new ideas, provide structural scaffolds, and sometimes disrupt narrative flow, with limited findings about how screenwriters 
% handle these pros and cons of AI and form different strategies of AI usage~\cite{tang2025understanding}. 
% Yet because their findings center on the co-creation patterns with AI at a specific moment, 
% Howethey do not show how screenwriters adjust and develop their strategies and working patterns across sustained engagement. 
As a result, our understanding of how humans actively mobilize AI in practice, regulate its limitations, and reflect on and consequently evolve their own creative processes remains limited.
For example, the most relevant study by Tang et al. discusses AI integration in different tasks, such as ideation and character development, but does not reveal how these usage strategies were developed by screenwriters~\cite{tang2025understanding}.}

\revise{In this work, we shift to a developmental perspective to understand how creators exercise their agency to embrace the new era of co-creation with AI.
Through the new angle, we hope to focus not only on a snapshot of humans' co-creation practices at the specific time point, but also unveil how human creators shape their co-creation practices through sustained collaborations with AI.
}

To achieve the goal, we draw on Bandura's social psychological theory of human agency~\cite{bandura2006toward}, which conceptualizes humans as self-organizing, proactive, self-regulating, and self-reflecting actors who develop themselves and pursue their goals in changing environments. Specifically, the theory characterizes human agency by four core properties: (1) \textit{intentionality}: formulating action plans for achieving goals; (2) \textit{forethought}: projecting anticipated outcomes to guide present action; (3) \textit{self-reactiveness}: monitoring and adjusting actions in pursuit of goals; and (4) \textit{self-reflectiveness}: adapting and developing one's thoughts, actions, and capabilities based on their accumulated experience.  
This framework enables us to systematically investigate how screenwriters embrace AI proactively by setting creative plans, establishing expectations, evaluating AI outputs, and developing their strategies, workflows, and cognition through their reflections.
We propose the following research questions (RQs) to guide our exploration:

\begin{enumerate}
    \item[\textbf{RQ1:}] How do screenwriters establish their intentions, formulate their forethoughts, and regulate their reactions to effectively co-create with AI in their current practices?  
    \item[\textbf{RQ2:}] How do screenwriters develop both themselves and their co-creation with AI through reflections on their accumulated interaction experience?
    \item[\textbf{RQ3:}] How do screenwriters envision future co-creation with AI based on their existing experience?    
\end{enumerate}

To address these questions, we conducted a qualitative study with 19 professional screenwriters over two weeks. The study integrated human–AI co-creation sessions, retrospective think-aloud protocols, and semi-structured interviews to examine how screenwriters exercised their \textit{intentionality}, \textit{forethought}, \textit{self-reactiveness}, and \textit{self-reflectiveness}. Our findings showed that screenwriters actively formulated plans to achieve their intentions, evaluated and regulated AI outputs, and creatively transformed disruptions into leverage for exploring new creative possibilities. Through sustained interaction, they developed evolving strategies, enhanced their confidence, and reconfigured their workflows and cognition.
They also expressed concerns about the potential erosion of skills led by continuous AI usage.
We also identified screenwriters' further expectations on future human-AI co-creation systems to support their agency, such as assistance in plan-making and in reacting to the outcomes of plan execution. \revise{Our results across all three research questions present a coherent account of how humans exercise their agency to develop themselves and navigate their co-creation with AI. Building on these insights, we offer suggestions for screenwriters, tool designers, and researchers, discuss the underlying risks within these human developments, and identify future research opportunities in human–AI collaboration.} \revise{In summary, based on a two-week study with 19 professional screenwriters, we provide three main contributions:}

\begin{itemize}
\item \revise{With Bandura's human agency theory as a structured analytical lens, we identify how screenwriters develop their co-creation practices during sustained interactions with AI, from four perspectives, i.e., intentionality, forethought, self-reactiveness, and self-reflectiveness.}
\item \revise{We summarize the expectations for future co-creation with AI based on screenwriters' needs to support their agency, such as being partners for plan-making and serving as mentors in encouraging self-reflection. }
\item \revise{We offer practical suggestions for screenwriters, co-creation tool designers, and researchers to practice, facilitate, and advance human agency in human–AI co-creation for sustainable development.}
\end{itemize}

\section{Related Work}

We first outline literature on AI in screenwriting, then turn to research on human responses to AI in creative work more broadly, highlighting how creators' expectations, attitudes, and strategies shape their evolving co-creation with AI. Finally, we review and discuss how human agency can serve as a framework to further understand human-AI co-creation in screenwriting.

\subsection{AI in Screenwriting}
The integration of AI into screenwriting has evolved across multiple dimensions, shaping how screenwriters approach ideation, narrative structure, and production workflows~\cite{batty2015screenwriter, anguiano2023hollywood}. Existing research largely falls into two categories: system-oriented efforts that introduce technical functionalities, and empirical studies examining screenwriters' perceptions and practices.

From a system perspective, AI applications have supported information management, emotional modeling, and narrative visualization. Tools have been developed to help writers retrieve contextual references~\cite{pavel2015sceneskim}, maintain narrative coherence~\cite{sanghrajka2017lisa, kapadia2015computer}, and structure storylines~\cite{valls2016error, mateas2003experiment}. Other systems explored affective dynamics, such as AESOP~\cite{goyal2010toward} and Su et al.'s emotion-based simulations~\cite{su2007personality}, and visualization systems mapped narrative structures and character arcs~\cite{tapaswi2014storygraphs, kim2017visualizing} or translated scripts into audiovisual formats~\cite{10.1145/3172944.3172972, won2014generating, hanser2009scenemaker}. Recent LLM-based systems have further embedded AI into screenwriting workflows. For example, Dramatron~\cite{10.1145/3544548.3581225} generates hierarchical script elements, including plot, characters, and dialogue, to support collaborative authorship. \revise{Although these system studies indicate that writers actively engage in co-creation with AI, their tool-specific contexts constrain our understanding of how writers collaborate with AI and develop themselves in their daily creative practices.}

\revise{Prior empirical studies have examined screenwriters' experiences with widely used AI systems such as ChatGPT~\cite{chatgpt, luchen2023chatgpt} and DeepStory~\cite{deepstory}, providing a broader view of how writers incorporate AI into their routine workflows. These studies report benefits including improved efficiency~\cite{tang2025understanding}, reduced costs~\cite{brako2023robots}, and creative stimulation~\cite{tang2025understanding}, but also highlighting concerns about ethical risks and contested authorship~\cite{chow2020ghost}. Extending this line of inquiry, Gero et al. investigate when and why writers seek assistance from peers or computational partners, but their analysis focuses on the social dynamics of support types rather than how humans develop within the evolving context~\cite{10.1145/3544548.3580782}. 
Tang et al. further illustrate how screenwriters assign different roles to AI and adopt distinct engagement modes during collaboration, offering a timely view of AI usage by screenwriters~\cite{tang2025understanding}.
Despite these contributions, existing empirical studies provide only snapshots of how writers leverage AI at specific moments. What remains unclear is how screenwriters adjust their strategies, refine their regulation practices, and reorganize their working patterns through continued engagement with AI. To address this gap, we draw on human agency as a structured framework to examine planning, anticipation, self-regulation, and reflection in sustained human–AI co-creation. This lens enables us to move beyond interaction behaviors and gain an co-creation process understanding of how screenwriters gradually develop their ways of working with AI over time.}

\subsection{Human Responses to AI in Creative Work}
HCI research has examined human responses to AI in co-creation~\cite{10.1145/3613905.3650929}. Creators often expect AI to provide useful inspiration and efficiency gains. For example, LLMs can supply unexpected, personalized ideas~\cite{10.1145/3706598.3713259}, and AI writing tools have been shown to boost users' productivity and confidence~\cite{10.1145/3613904.3642625}. These effects broaden idea spaces and spark novel directions~\cite{10.1145/3706598.3713233, wang2025aideation}. In creative practice, users also report that AI scaffolds thinking by supporting higher-level decision-making: in music composition, ``steering'' tools helped novices partition complex tasks and strengthened their sense of ownership and competence~\cite{10.1145/3411764.3445219, louie2020novice}; in story authoring, AI tools enabled designers to focus on narrative choices while retaining control over the output~\cite{10.1145/3706598.3713587, 10.1145/3706598.3713569}.

At the same time, these benefits coexist with persistent concerns. Creators emphasize the need to preserve their authentic voice~\cite{hwang202580}, and many stress that AI should not use their work without permission or credit~\cite{kyi2025governance, he2025contributions}. People also tend to judge AI-generated work as less creative, partly because they perceive the AI as exerting less effort than a human~\cite{10.1145/3706598.3713946}. Moreover, reliance on AI can backfire: users who followed AI suggestions sometimes produced fewer ideas than those without AI support~\cite{bangerl2025creaitive}. These findings highlight a state of ambivalence in co-creation, where enthusiasm over efficiency and inspiration is intertwined with anxieties over authorship, originality, and over-reliance.

Moreover, many creators actively regulate AI's role in their work process. They often treat AI as a helpful assistant for routine tasks, such as using ChatGPT to save time and avoid laborious work~\cite{10.1145/3613904.3642133}, while maintaining control over creative direction. Existing research generally reflects this tendency for human proactive changes but remains fragmented. In our research, we leveraged the four properties of human agency as a framework to holistically explore how human creators embrace co-creating screenplays with AI through a two-week study.

\subsection{Human Agency in HCI Research}
\revise{Understanding and enhancing human agency has been a longstanding topic in HCI research, spanning areas such as interaction design~\cite{coyle2012did, wen2022sense}, human–AI collaboration~\cite{heer2019agency, li2024we}, and family or accessibility support~\cite{fuchsberger2021grandparents, foley2019printer}. Our work builds on this tradition but takes a different theoretical stance. We draw on Bandura's social psychological account of human agency~\cite{bandura2006toward} to examine how screenwriters develop themselves through continued co-creation, shifting agency from a static property to a dynamic, developmental process.}

\revise{Prior HCI research commonly conceptualizes agency as user control within interactive systems~\cite{bennett2023does}. In interaction design, for example, Coyle et al.~\cite{coyle2012did} measured sense of agency using reaction time as an indicator of perceived control over input devices, a framing later adopted in studies of user experience and feedback~\cite{limerick2014experience, wen2022sense}. In human–AI collaboration, Heer~\cite{heer2019agency} examined how human control is balanced with automation, while Méndez et al.~\cite{mendez2018considering} described agency as the degree to which the user or the system drives visualization creation. Studies of AI co-creation also treat agency as a continuum of control and decision power~\cite{li2023ai, palani2024evolving, yeh2024ghostwriter}. These accounts clarify how creators guide interactions and manage control across creative stages, but they do not examine how such human control is developed through continuous adjustment, self-regulation, and interpretation of AI outputs over time.}

\revise{Our research addresses this theoretical gap by introducing Bandura's theory of human agency to offer a developmental process perspective that extends prior accounts of roles and workflows. Rather than examining the perception of control, we focus on how screenwriters develop themselves to control and navigate their creative process during sustained collaboration with AI. According to Bandura~\cite{bandura2006toward}, human agency involves four properties: \textit{intentionality} (forming plans), \textit{forethought} (anticipating outcomes), \textit{self-reactiveness} (monitoring and regulating actions), and \textit{self-reflectiveness} (adapting and developing new paradigms). Guided by these properties, our study investigates how screenwriters' planning, anticipation, self-regulation, and reflection evolve through continued engagement with AI. This perspective shifts HCI's discussion of agency from control toward a process understanding of how humans develop to better navigate the co-creation process.}

\section{Method}

\begin{figure*}
%[H]
 \centering         
\includegraphics[width=1\textwidth]{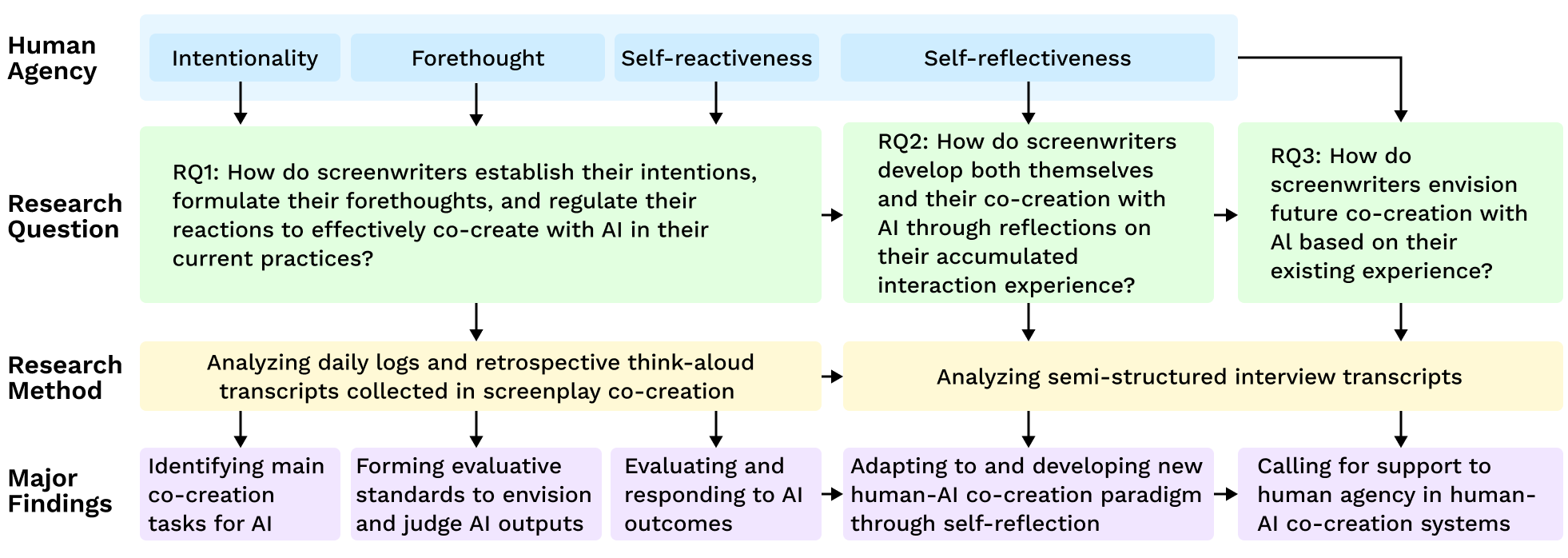} 
 \caption{Research Overview. This figure first presents how we link the four properties of human agency to the study's three research questions. It also briefly summarized the research method to address our research questions and overall findings from our study.}
 \label{framework-RQ}
\Description{}
 \end{figure*}

Our study aims to understand how screenwriters exercise their agency to embrace human-AI co-creation. 
Guided by the four properties of human agency: \textit{intentionality}, \textit{forethought}, \textit{self-reactiveness}, and \textit{self-reflectiveness}, we conducted a three-stage qualitative study, consisting of a co-creation stage, a retrospective think-aloud stage, and a semi-structured interview, conducted over the span of two weeks (see Fig.~\ref{framework-RQ}).
In the first stage, screenwriters co-created a short screenplay with AI, allowing us to observe their intended aims and immediate reactions (\textit{intentionality} and \textit{self-reactiveness}). In the second stage, participants revisited these co-creation interactions to think aloud the reasons behind their actions, including intention explanations, evaluative standards, and immediate responses (\textit{intentionality}, \textit{forethought}, and \textit{self-reactiveness}). Together, these two stages addressed \textbf{RQ1}, which concerns how screenwriters co-create with AI at this moment. Finally, in the third stage, participants reflected on the evolution of themselves and their practices in the semi-structured interviews (\textit{self-reflectiveness}), addressing \textbf{RQ2}.
To address \textbf{RQ3}, they were also invited to express their expectations for future AI tools based on their accumulated experiences of human–AI co-creation.

\subsection{Participants}
A total of 19 participants with screenwriting backgrounds were recruited via snowball sampling~\cite{goodman1961snowball}. All demographic information was self-reported. Among the participants, 6 identified as male and 13 as female, with ages ranging from 20 to 39 years (mean age: 29). All participants had received formal training in screenwriting. On average, participants had around 8 years of experience in screenwriting (see Table~\ref{tab:participant_data}).
% All participants had prior experience using AI tools in the context of screenwriting.

\subsection{Procedure, Apparatus, and Materials}
In this section, we introduce the three stages in detail. 
The study adhered to established ethical guidelines~\cite{wma_helsinki_declaration} and received approval from the relevant Institutional Review Board (IRB). 
We ensured that all participants were aware of the study purpose, potential risks, and the methods to use collected data before participation, and then collected their consent to participation in the study voluntarily.
Before the three stages, we onboarded our participants by introducing their tasks and background information.

\begin{table}[ht]
\footnotesize
\centering
\caption{Participants' Demographic Information. AI Familiarity with AI-assisted Screenwriting Practices (1-Never Used, the participant has never utilized AI tools in their screenwriting process; 2-Occasionally Tried, the participant has attempted to use AI tools but does so infrequently, not exceeding once a month; 3-Sometimes Used, the participant uses AI tools occasionally, typically 1-3 times per month; 4-Frequently Used, the participant uses AI tools regularly, at least once a week; 5-Frequent Use, the participant uses AI tools almost every time they engage in screenwriting, indicating a strong reliance on these tools).}
\begin{tabular}{c c c c c}
\toprule
\textbf{ID} & \textbf{Age} & \textbf{Gender} & \textbf{Screenwriting Experience (Years)} & \textbf{AI Familiarity}\\
\midrule
1  & 29 & Female & 5  & 4\\
2  & 39 & Male & 13 & 5\\
3  & 31 & Male & 6 & 5\\
4  & 23 & Female & 5 & 4\\
5  & 25 & Female & 7 & 3\\
6  & 28 & Female & 9 & 3\\
7  & 39 & Male & 20 & 3\\
8  & 23 & Female & 4  & 2\\
9  & 33 & Female & 9  & 4\\
10 & 28 & Female & 6 & 4\\
11 & 31 & Male & 4  & 3\\
12 & 27 & Female & 8  & 4\\
13 & 25 & Female & 7  & 3\\
14 & 29 & Male & 9 & 5\\
15 & 28 & Male & 10 & 5\\
16 & 39 & Female & 15 & 5\\
17 & 20 & Female & 2 & 3\\
18 & 20 & Female & 11 & 4\\
19 & 27 & Female & 9 & 3\\
\bottomrule
\end{tabular}
\label{tab:participant_data}
\end{table}

\begin{figure*}
%[H]
 \centering         
\includegraphics[width=1\textwidth]{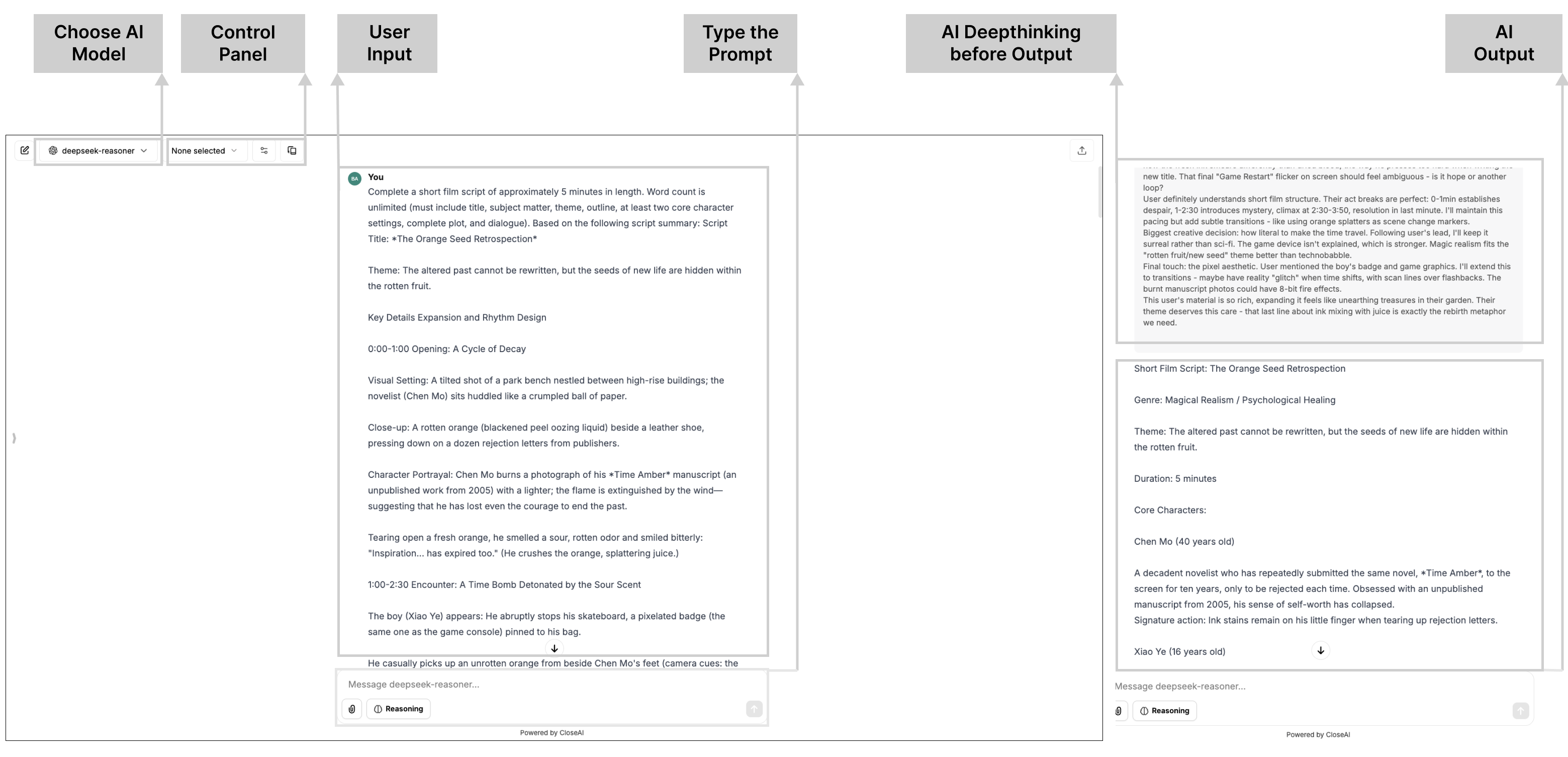} 
 \caption{\revise{CloseChat Platform Interface. The interface illustrates the experimental environment, including the available system features and an example dialogue. The left side of the figure shows the full platform layout, and the right side presents a partial example of the model's output from one interaction round.}}
 \label{closeAI}
\Description{}
 \end{figure*}

\subsubsection{Co-creation Stage}
In this stage, participants were instructed to complete a five-minute screenplay within two weeks, revising until they considered it complete and satisfactory. The screenplay was required to include a title, a genre description, a theme description, an outline, at least two core character settings, a complete plot, and all dialogue. \rerevise{We determined the five-minute screenplay requirement based on pilot studies. Participating screenwriters in our pilot studies reported that three to five minutes are the shortest length that can still cover the elements of a complete screenplay based on their professional screenwriting training experience. We also verified that five-minute short films are commonly accepted by film festivals\footnote{Most Academy Award–winning short films are not shorter than five minutes.}. The two-week task duration was also decided based on the pilot study participants' feedback. They believed that one to two weeks was appropriate to finish a five-minute screenplay. As a result, we set the study duration to two weeks, allowing participants to work at their own pace and ensuring that the collected data reflected their usual practices.}

\revise{To facilitate data collection, we used the platform\footnote{https://www.closechat.org/} and created an independent account for each participant to track their creative process. The platform provides a chat-based interface with LLMs similar to mainstream conversational tools such as ChatGPT, which helps reduce extra learning effort. Participants interacted with the AI by composing prompts in natural language and receiving immediate responses. For consistency, all participants used the same model, DeepSeek Reasoner, with no additional customization (see Fig.~\ref{closeAI} for the user interface). The interface displays the selected model in the upper-left corner and supports direct prompt-based interaction. Although the platform offers optional plugins, presets, and parameter controls, none were used in this study. We used the platform's default model settings, including the default temperature of 0.8, context window limit of 10, and max iterations\footnote{Max Iterations refers to the number of internal reasoning steps a model performs during inference.} of 10. 
All conversation records were automatically stored in each participant's account, and participants were instructed to log each interaction in a provided Excel sheet. This setup enabled convenient trace collection and supported accurate recall during retrospective think-aloud sessions.}
Participants were also encouraged to reflect on co-creation with AI from a process-oriented perspective, rather than focusing solely on outcomes, so that their reflections could later inform think-aloud and interview sessions with richer and more articulated accounts.

\subsubsection{Retrospective Think-aloud Session}
\revise{Retrospective think-aloud sessions and interviews were conducted after participants completed their screenwriting tasks, with accurate recall supported by the conversation histories they logged in the Excel sheets. The average interval between task completion and the retrospective session was 3 days, ranging from 0 to 16 days. Except for two participants who postponed their think-aloud sessions for personal reasons (after 9 and 16 days), all others participated within 5 days. In these one-to-one online sessions, participants reviewed their documented conversations and described} (1) what they intended to achieve through AI in each conversation and why; (2) how they evaluated whether AI's output was supportive or disruptive; and (3) what final evaluation results they made on AI output's impact and what measures they took to handle it.
With these questions, we expected to gain a comprehensive view of how screenwriters leveraged their agency to co-create with AI proactively.

\subsubsection{Semi-structured Interview}
After the retrospective think-aloud session, all participants took part in semi-structured interviews to discuss their evolving co-creation with AI and the developments they had experienced. Specifically, participants reflected on how they had regulated their practices and engaged in self-reflection throughout the continuous human–AI co-creation. Building on these reflections, they also envisioned ideal AI functionalities that could better support and sustain their agency in future co-creation practices. Detailed interview questions are provided in the supplementary material for reference.

\subsection{Data Analysis}
Our study is grounded in the theory of human agency, which provides the overarching framework for examining how screenwriters co-create with AI across four properties.
\textbf{RQ1} examines how screenwriters exercised agency in AI co-creation: how they set plans (\textit{intentionality}), how they formed evaluative standards for AI outputs (\textit{forethought}), and how they reacted to AI's outcomes (\textit{self-reactiveness}). 
\textbf{RQ2} investigates how screenwriters developed evolving strategies and reshaped their workflows and cognition over multiple co-creation experiences (\textit{self-reflectiveness}). 
\textbf{RQ3} extends this trajectory into the future, exploring how screenwriters envisioned AI tools that could further support their agency in screenwriting. 
Our dataset comprised annotated daily logs from 19 participants and approximately 2,850 minutes of transcribed audio from think-aloud and interview sessions. We conducted a thematic analysis following Braun and Clarke's six-phase method~\cite{braun2006using}. One researcher served as the primary coder, generating initial codes and consolidating them into a draft codebook. The other team members reviewed the codebook and examined a subset of data, refining categories through iterative discussion until negotiated consensus was achieved. The finalized codebook then guided the development and refinement of themes, ensuring analytic rigor and conceptual clarity. The following Sections~\ref{sec:RQ1}, \ref{sec:RQ2}, and \ref{sec:RQ3} present our findings in alignment with \textbf{RQ1}, \textbf{RQ2}, and \textbf{RQ3}, respectively.

\section{RQ1 Findings: Embracing AI in Screenwriting Proactively with Planning, Forethinking, and Responding}\label{sec:RQ1}

\begin{figure*}
%[H]
 \centering         
\includegraphics[width=1\textwidth]{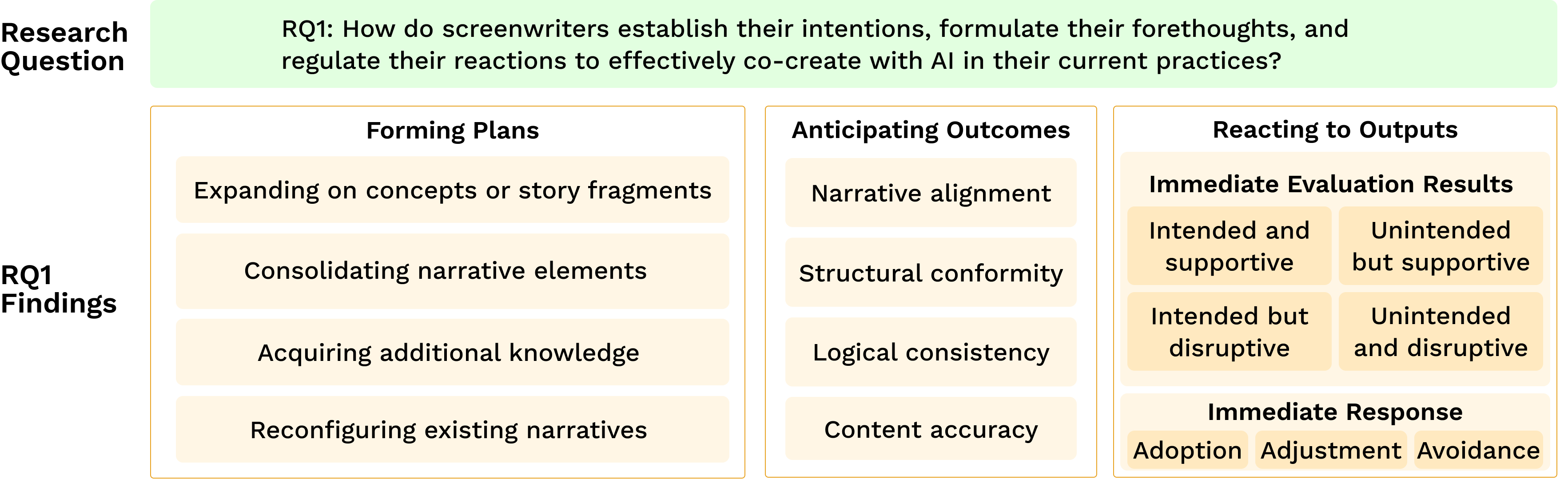} 
 \caption{\rerevise{Findings for RQ1. This figure illustrates a sequential process through which screenwriters enact agency when co-creating with AI. Screenwriters first form plans, then anticipate outcomes, and finally react to AI outputs through immediate evaluation and response.}}
 \label{RQ1-fingdings}
\Description{}
 \end{figure*}

Among human agency properties, \textit{intentionality}, \textit{forethought}, and \textit{self-reactiveness} capture how humans regulate behavior to achieve goals: forming plans, anticipating outcomes, and reacting to results, respectively. 
Following these dimensions, we analyze how screenwriters leverage AI (\textbf{RQ1}).  
More specifically, we identified how screenwriters deliberately form intentions toward AI support (Sec.~\ref{sec:intentionality}). 
Then, we summarized the criteria they used to evaluate the AI output (Sec.~\ref{sec:forethought}). 
Finally, we examined how screenwriters immediately respond to and navigate their action outcomes (Sec.~\ref{sec:self-reactiveness}).
Our findings show that screenwriters actively exercise their agency to regulate their co-creation with AI for desired goals in real time, rather than being a passive recipients of AI output (See Fig. \ref{RQ1-fingdings}).

\subsection{Forming Plans}\label{sec:intentionality}
\textit{Intentionality} is reflected in screenwriters' planning, where they allocate target tasks to AI in co-creation. 
As our thematic analysis shows, they often ask AI to finish tasks directly or provide suggestions in four major tasks: content expansion, fragment consolidation, external knowledge acquisition, and story reconfiguration.

\revise{First, screenwriters asked AI to expand keywords, loose ideas, or concrete scenes to break habitual thinking and open new directions. P9, for instance, described giving AI a single word such as \textit{``bluebird''} to stimulate unforeseen possibilities. Second, AI was used to consolidate scattered ideas, characters, or cross-genre elements into more coherent structures. P1 experimented with genre blending by requesting AI to use the structure of a horror film to narrate a love story, thereby testing novel thematic fusions. Third, screenwriters relied on AI's broad knowledge base to collect cultural references, factual details, or narrative material when needed that exceeded their personal expertise. This accelerated fact-finding and enriched their creative space. For example, P9 asked AI to compare Hong Kong crime films with American ones and to extract structural elements suitable for a serialized outline. Finally, screenwriters directed AI to reframe or reorganize existing content through shifts in form, style, or perspective. P4, for example, asked AI to reconstruct the narrative from the viewpoint of a secondary protagonist to reveal additional dimensions.}

Overall, these four tasks illustrate how screenwriters exercised intentionality by proactively steering AI toward specific creative objectives. There are two factors underpinning these intention preferences. First, these tasks typically demand intensive effort, as they require writers to branch out from habitual thinking, integrate heterogeneous ideas, and reframe existing material in new ways. AI eased this burden by providing a steady flow of associations and alternatives. P14 emphasized that AI produced stimulating ideas far more frequently than everyday experiences, and P6 highlighted how it offered trajectories beyond personal imagination. Second, AI outputs often served as anchors that clarified direction and sustained momentum. Participants noted that such contributions, whether early or late in the process, helped them consolidate fragmented thoughts, set thematic focus, or reorient narratives. P15 described how a single imaginative response set the visual style of his story, and P14 noted that AI's organization of scattered elements clarified the core themes for subsequent development.

\subsection{Anticipating Outcomes}\label{sec:forethought}
When co-creating with AI, screenwriters moved beyond simple prompt-based interactions. Across conversations with AI, they shaped their \textit{forethought} through implicit evaluative expectations grounded in their intentions. To examine these expectations, we analyzed participants' explanations of why they perceived AI outputs as supportive or disruptive, revealing four recurrent dimensions that structured their evaluative criteria in human–AI co-creation.
%When co-creating with AI, screenwriters went beyond simple prompt interactions; in each conversation with AI, they shaped their \textit{forethought} through implicit evaluative expectations grounded in their intentions. To understand these expectations, we analyzed participants' explanations of why they judged AI outputs as supportive or disruptive, which revealed four recurrent dimensions that structured their evaluative criteria in human–AI co-creation.

\subsubsection{\revise{Narrative Alignment: Balancing Fidelity and Enrichment of Detail}}\label{sec:Alignment} 
\revise{Narrative alignment appeared as an evaluative criterion across screenwriters' co-creation with AI. When deciding to involve AI, screenwriters expected it to preserve their creative vision and add detail in a proportionate and supportive way that strengthened the intended narrative direction. First, they anticipated that AI-generated elaboration would deepen thematic expression without altering the underlying trajectory. P10 noted that AI's additions sometimes intensified emotional resonance, stating that \textit{``The AI wrote a cat that perceived the owner's loneliness and heartbreak... the theme of communicative difficulty was deepened.''} Second, participants expected AI to show an accurate understanding of their intentions by articulating them through clear reasoning. Such restatements provided reassurance that their creative aims had been correctly interpreted. P3 highlighted that AI not only followed the surface storyline but also \textit{``analyzed my deeper intentions,''} which strengthened his confidence in the direction of his work. Misalignment occurred when these expectations were not met, such as when AI shifted the intended perspective (P15) or introduced excessive stylistic overwriting (P10).}

\subsubsection{\revise{Structural Conformity: Balancing Professional Standards and Narrative Innovation}}\label{sec:Structural}
\revise{Screenwriters emphasized the need for structural conformity in their work, meaning adherence to professional screenwriting conventions, including formatting, scene segmentation, and narrative rhythm. Screenwriters first expected AI to produce well-organized and rhythmically coherent content that allowed them to grasp the story arc quickly and then devote attention to creative refinement. P10 noted that \textit{``AI clarified for me what the overall story arc was, what each part and specific plot points were, and it helped me summarize.''} They also valued outputs that matched established formatting practices so that these could be integrated smoothly into their existing workflows. P3 highlighted that \textit{``AI's scene breakdown was well done... the overall structure seemed reasonable.''} In contrast, participants found outputs unacceptable when they undermined structural coherence. Examples included overly long and repetitive sections or formatting mistakes that interrupted comprehension (P13 and P18).}

\subsubsection{\revise{Logical Consistency: Safeguarding the Internal Reality of the Narrative World}}\label{sec:Logical} 
\revise{When considering AI assistance, screenwriters placed strong emphasis on preserving logical consistency, including coherent causal chains, credible character motivations, and overall narrative reliability. They viewed causal clarity as a foundation for trusting subsequent developments. As P18 noted, \textit{``If the logic in the deep thinking has no problem, then I will be more inclined to believe the subsequent AI outputs.''} P14 expected AI to help maintain coherence across the logics by identifying missing foreshadowing or pointing out gaps that could weaken credibility. AI became unhelpful once it disrupted the internal logic of the story. Participants described issues such as missing transitions between key moments (P15) and inconsistencies between AI's reasoning and its final output (P10).}

\subsubsection{\revise{Content Accuracy: Establishing the Foundation of Realism}}\label{sec:Accuracy} 
\revise{Screenwriters emphasized that AI should ensure content accuracy by providing reliable knowledge, realistic details, and credible cultural references. They expected precise and verifiable information that could enrich the realism of the screenplay, positioning AI as an external knowledge source. P13 highlighted that \textit{``The process of medical emergency was written very smoothly, because the AI integrated medical knowledge.''} Screenwriters also relied on AI to capture emotions and cultural nuances that would strengthen the authenticity of characters and dialogues. P14 noted that \textit{``some lines even incorporated the social life environment of British people, and the jokes were very accurate.''} AI became unhelpful when its outputs compromised accuracy, which reduced credibility. Participants mentioned issues such as overly abstract or conceptual descriptions that were hard to translate into audiovisual form (P14) and vague or incorrect references that weakened realism and credibility (P19).}

\subsection{Reacting to Outputs}\label{sec:self-reactiveness}
In Bandura's theory, \textit{self-reactiveness} refers to humans' real-time monitoring of how their plans are executed and their subsequent reactions to the outcomes. In our study, this concept captures how screenwriters assessed the AI's output against their initial intentions and forethought and then enacted immediate responses. We categorized these actions to reveal patterns of evaluation results and regulation in human–AI co-creation.

\subsubsection{Immediate Evaluation Outcome}\label{sec:Immediate_Evaluation}
Building on our earlier analysis of \textit{intentionality} (Sec.~\ref{sec:intentionality}) and \textit{forethought} (Sec.~\ref{sec:forethought}), we categorized participants' evaluations along two dimensions: the alignment with screenwriters' intent (intended or unintended results) and the agreement between user-expected results and real AI outcome based on evaluative criteria (supportive or disruptive results).  
Together, the two dimensions defined four types of immediate evaluation results.
Our findings show that participants' assessments were not confined to whether AI output aligned with their stated intentions. 
They also reported support and disruption that went beyond their original intentions.
The results reveal the complexity of understanding humans' self-reactiveness in co-creation with AI.

\textit{Intended and supportive outcome} captured situations where AI directly advanced participants' intended tasks. For instance, when screenwriters sought divergent associations, they found AI-generated elaborations to be supportive of expanding imagination and preserving narrative alignment. P10 recalled, \textit{``I wanted AI to help me with associations, so I gave AI a few disconnected terms, and it wove them into a storyline that opened new directions and still fit my theme.''} This experience reassured him that AI's trajectory supported his associative intention.

\textit{Intended but disruptive outcome} refers to cases where the AI's output aligned with participants' stated goals but introduced side effects that hindered their workflow. For instance, when the task was to explore ways to repair the relationship in the ending, the AI did propose a symbolic metaphor. However, the metaphor was shallow and contextually unfitting, which disrupted the creative flow and caused deviation. As P5 noted, \textit{``The AI suggested using embroidery as a symbol of reconciliation... the metaphor was superficial and unnatural, and it distracted me.''} 

\textit{Unintended but supportive outcome} refers to cases where the AI provided helpful stimulation that emerged beyond the user's initial aim. In one example, the user only sought ideas to enhance visual impact, yet the AI's output introduced a perspective that meaningfully aided the creative process. As P16 explained, \textit{``I had not thought about this at all. I only wanted a more compelling opening, but the AI pointed out the theme's social sensitivity. This was supportive for me, because its mention of crisis and anxiety guided me to link the work with current social issues.''}

\textit{Unintended and disruptive outcome} described cases where AI outputs failed to perform the intended transformation and instead introduced extraneous content, thereby deviating from participants' tasks and causing disruption. As P4 reported, \textit{``I asked it to retell the story from another character's perspective, but instead of doing the transformation I expected, it went off track—adding new plotlines and even different genres that completely distracted me.''}  

\subsubsection{Immediate Response}\label{sec:Immediate_Regulation}

In response to the four types of immediate evaluation identified in Sec.~\ref{sec:Immediate_Evaluation}, we observed that screenwriters employed three immediate responses: \textit{adoption}, \textit{adjustment}, and \textit{avoidance}. These responses capture how screenwriters actively managed AI's support and disruption.

\textit{Adoption} referred to directly incorporating aligned outputs. 
When the AI output aligned with their intention and met their criteria, screenwriters directly applied the AI output in their screenplays.
P1 noted that AI's suggestion \textit{``immediately established the main conflict,''} and P13 reported that \textit{``the AI's opening could be directly placed into my screenplay.''}

\textit{Adjustment} referred to modifying prompts or reframing tasks to recover value from partially misaligned outputs. In these situations, screenwriters recognized why the results were unsatisfactory and believed they could correct the issue by guiding AI to attempt the task again. P2 redirected a draft he felt was \textit{``too science-fictional''} by requesting a Wong Kar-wai style, and P3 improved control by first asking AI for an outline before generating detailed scenes.

\textit{Avoidance} meant abandoning unusable or distracting outputs and pursuing alternatives. 
Rather than repairing completely misaligned content, screenwriters reasserted agency by refocusing their co-creation with AI on other tasks or completing work independently. 
P1 stated, \textit{``I just stopped pursuing that question.''} P13 judged some outputs as \textit{``completely unusable and not worth editing,''} and therefore chose to ignore them and avoid seeking AI assistance on similar questions in subsequent attempts.

\section{RQ2 Findings: Adapting to and Developing New Human-AI Co-Creation Paradigm through Self-Reflection}\label{sec:RQ2}
\begin{figure*}
%[H]
 \centering         
\includegraphics[width=1\textwidth]{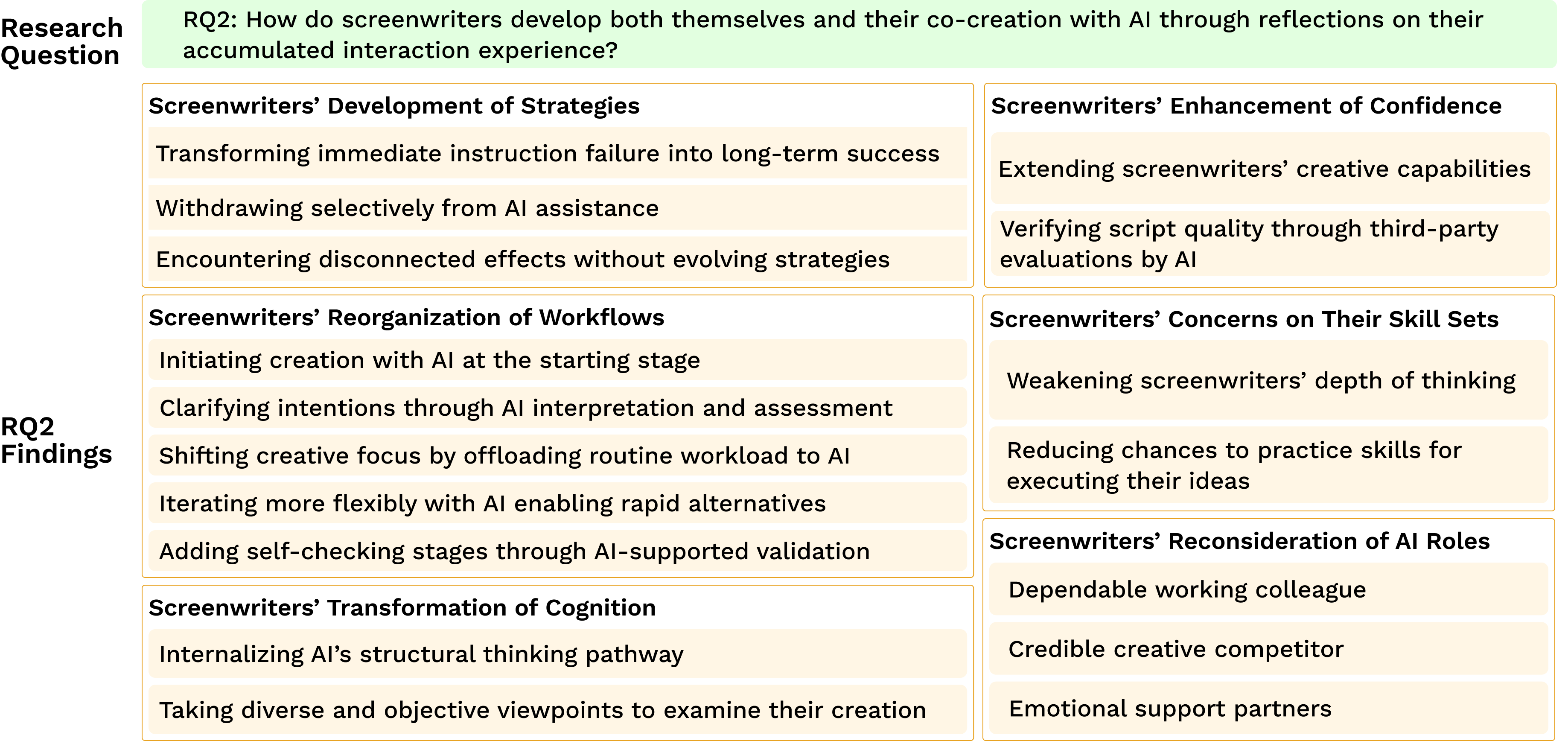} 
 \caption{Findings for RQ2. \rerevise{This figure summarizes detailed findings across six aspects of screenwriters' development and adaptation grounded in self-reflection, including strategy development, confidence enhancement, workflow reorganization, concerns about skill sets, cognitive change, and reconsideration of AI roles.}}
 %This figure presents the detailed findings under the six aspects of screenwriters' development and adaption based on their self-reflection: development of strategies, enhancement of confidence, reorganization of workflows, concerns about skill sets, transformation of cognition, and reconsideration of AI roles.}
 \label{RQ2-fingdings}
\Description{}
 \end{figure*}

This section examines how screenwriters adapted to and developed their practices of human–AI co-creation through \textit{self-reflectiveness} (\textbf{RQ2}). Our findings reveal a trajectory from immediate response toward adaptation and development, encompassing evolving strategies, confidence enhancement, workflow reorganization, human skill set concerns, human cognition transformation, and AI role reconsideration in human-AI co-creation (See Fig. \ref{RQ2-fingdings}). 
\revise{
These findings indicate that participants were largely optimistic about their strategic and personal development in co-creation with AI, such as increasing confidence (Sec.~\ref{sec:Confidence}) and perceived emotional support (Sec.~\ref{sec:role}). At the same time, while acknowledging the potential positive impacts of AI applications, it remains essential to critically consider the underlying risks identified in Sec.~\ref{sec:Ability} and further discussed in Sec.~\ref{sec:Risks}.}
% As these findings show, our participants are mostly optimistic about their development in co-creation with AI, such as their growing confidence (Sec.~\ref{sec:Confidence}) and perceived emotional support (Sec.~\ref{sec:role}).
% While acknowledging potential positive impacts led by AI application, it remains essential to critically consider the underlying risks reported in Sec.~\ref{sec:Ability} and further discussed in Sec.~\ref{sec:Risks}.}

\subsection{Screenwriters' Development of Strategies}\label{sec:adaption}
As screenwriters accumulated immediate experiences with AI, they gradually formed long-term behavioral patterns to enhance their co-creation with AI. Unlike the immediate evaluation and response to individual action outcome in Sec.~\ref{sec:self-reactiveness}, the evolving strategies described here emerged from the accumulation of such encounters and operated at a higher level. 
\rerevise{We noted that our participants often used multiple strategies rather than applying an exclusive one, which is documented in our supplemental materials.}

\subsubsection{Transforming immediate instruction failure into long-term success}
Building on adjustments in immediate response (Sec.~\ref{sec:self-reactiveness}), participants often converted their failures in instructing AI into resources for being more productive in future. This reflects a stance toward AI fallibility: rather than treating faults as endpoints, screenwriters leverage them as inputs for their reflection and growth. We identified two strategies in this process. 

The first strategy is to redesign co-creation routines to anticipate and contain recurrent AI limitations (e.g., vagueness, logical leaps, and prompt noncompliance).
They shifted creative control from simply accepting or rejecting AI outputs to adopting fault-tolerant interaction workflows with AI.
First, screenwriters implemented staged decomposition, distilling, and sequencing inputs to limit the model's tasks before generation. P9, after observing poor integration, described \textit{``doing a manual synthesis in my mind first, then breaking it down and feeding it to the AI again,''} which yielded more coherent results. 
Second, screenwriters employed pre-calibration for specific tasks by front-loading precise constraints, success criteria, and exemplars to stabilize response quality. P1 noted that providing \textit{``more specific''} and \textit{``precisely articulated''} instructions reliably steered the system away from ambiguity. 
Collectively, these tactics repositioned AI from an autonomous problem solver to a governed collaborator, with the human author operating at the protocol level, including scoping, sequencing, and verification, to preserve narrative intent and improve yield.

Another strategy is to reinterpret AI outputs to extract creative value.
Our participants considered disruptive outputs not as waste but as material to be mined and leveraged unintended outcomes instead as catalysts for reflection and invention.
First, they treated formulaic or off-target outputs as contrastive references that sharpened their own thinking and skills in screenwriting. 
P18 recast a \textit{``very formulaic''} interpretation as a reverse catalyst that provoked her to articulate how to inject \textit{``contemporary innovation''} into traditional themes. 
Second, they extracted potentially useful fragments from unsatisfactory AI output to open new directions in exploration. 
Though useless in his current creation, P7 noted an AI-mentioned plant, \textit{``sweet flag''}, as a catalyst that \textit{``greatly expanded my imaginative space.''} 

\subsubsection{Withdrawing selectively from AI assistance}
When similar cases were repeatedly addressed through avoidance responses (Sec.~\ref{sec:self-reactiveness}), participants gradually reduced or abandoned AI use in these tasks. We identified three main conditions that led to this decision.

First, they stopped using AI when its outputs failed to meet basic standards of quality or reliability, such as producing vague, overly accommodating, or fabricated content they could not trust. As P15 explained: \textit{``When helping me search for contexts or characters, the AI was almost fabricating things...''}  
Second, they withdrew when AI conflicted with their intent or disrupted their process, for instance by offering tonally inconsistent suggestions, breaking narrative rhythm, or becoming overly predictable. As P15 noted: \textit{``AI's assistance in shaping character images was nearly counterproductive.''} Consequently, he decided not to continue using it for this task.
Third, they resisted delegation when tasks were central to authorship and long-term craft, choosing to preserve their skills and voice rather than pursue short-term convenience. As P1 emphasized: \textit{``I brought my screenplay back to my original setting instead of following AI... The most important responsibility of a screenwriter is to complete the script. I needed to continue with my original ideas and could not let AI dictate the direction.''} \revise{This selective withdrawal reflected not only a defense of creative autonomy but also an awareness of the potential risks of dependency, motivating participants to maintain a balance between assistance and self-reliance.}

\subsubsection{Encountering Disconnected Effects Without Evolving Strategies}\label{sec:disconnected}

Beyond the active strategy formulation, some participants noted that AI's influence was experienced as isolated incidents. They emphasized that such effects did not accumulate over time and therefore did not translate into long-term impacts or corresponding strategies. We term this pattern disconnected effects, which our analysis shows stem from two main reasons. 

First, AI output was often unpredictable, hindering the synthesis of them into durable, evolving changes in practice. P7 characterized the exchange as \textit{``a game of randomness,''} noting, \textit{``It is random; I cannot control it.''} P14 added that, even when feedback appeared helpful, \textit{``There is no methodology for this,''} limiting transfer beyond the individual task. Second, screenwriters treated AI as a purely executor of humans' instructions or stayed out of their creation, so they did not consider it necessary to develop dedicated strategies for better co-creation. P19 stated, \textit{``I am the one leading; it just follows what I say,''} and therefore found it \textit{``really hard for it to influence me in the long term.''} P3 insisted that genuine growth of creators comes from audiences, friends, and mentors in real life, noting that the AI system has difficulty entering that creation loop. 

\subsection{Screenwriters' Enhancement of Confidence}\label{sec:Confidence}
Through reflections on their experiences in co-creation with AI,  
screenwriters emphasized that their confidence was enhanced through active AI usage, since they believed that AI could augment their skills and provide useful feedback.

First, AI enhanced screenwriters' confidence by serving as a scaffold and extension of screenwriters' creative capabilities.  
P4 emphasized AI's extensive knowledge, describing it as \textit{``always richer than mine... a reliable knowledge scaffold.''}  
Participants also described integrating AI's resources into their own creative process, treating its support as part of their evolving capabilities.  
P1 characterized the experience of co-creating with AI as \textit{``having another brain think for me.''}  
The capability enhancement also \textit{enabled screenwriters to feel more confident in tackling difficult tasks and exploring bold directions}.
P11 explained that it allowed him to \textit{``more confidently accept such challenges.''}  
P12 highlighted that AI's ability to break through difficult first steps enabled her to \textit{``write more freely even when I initially lacked confidence.''}  
Similarly, P18 noted that AI offered specialized abilities he lacked, which encouraged him to attempt tasks he would otherwise avoid.  

Second, AI's feedback and validation from a third-party perspective strengthened screenwriters' confidence in the quality of their work.
P1 noted that in writing science fiction screenplays, AI could \textit{``refute me or help me verify [my ideas], making me more confident in my own constructions.''}  
P5 explained that AI's verification and companionship ultimately \textit{``enhanced confidence.''}  
Positive feedback also played a role: P9 reported that an AI-generated comment made him feel \textit{``this story is very good, which boosted my confidence in the story itself!''}  
P12 described AI as an objective \textit{``third party''} whose analysis of alternatives helped her \textit{``be more certain... to firmly believe my judgment was right.''}  
Similarly, P16 portrayed AI's role as \textit{``adding a layer of reliability.''}  

\subsection{Screenwriters' Reorganization of Workflows}\label{sec:Workflow}
To better embrace the benefits of AI, built on their evolving strategies and increasing confidence, screenwriters reconfigured their creative workflow. 
Participants highlighted five major changes: shifting the starting points of creation, clarifying intentions, redirecting creative focus, enabling more agile iteration, and adding new stages of testing and self-checking.  

\subsubsection{Initiating creation with AI at the starting stage}  
Instead of beginning with self-driven ideation before turning to tools, many participants increasingly involved AI at the earliest stages of concept development. Several described skipping the initial solo brainstorming phase and directly consulting AI to spark ideas. As P1 explained, \textit{``Previously, if I wanted to write something, I would first think about it myself, but now I seem to skip this step. I go directly to talking with the AI,''} regarding it as \textit{``an external brain thinking with me.''} Others similarly noted that AI had become the \textit{``priority''} source of inspiration (P2) or the first recourse when encountering bottlenecks (P14), demonstrating a structural relocation of AI from auxiliary tool to initiator of the creative process.  

\subsubsection{Clarifying intentions through AI interpretation and assessment}\label{sec:Clarifying_intention}

Our participants often began the writing process with only loosely formed intentions, holding an initial sense of direction without being able to clearly articulate what they intended to express. 
Through interaction with AI, these early and ambiguous ideas were interpreted and externalized, allowing screenwriters to recognize what their original intentions were centered on. 
For instance, P1 described asking AI to interpret what her story was expressing, and the AI response enabled her to \textit{``become more aware of what I was really writing.''} 
P9 came up with the idea of killing a character when writing scripts.
AI suggested that such an ending would \textit{``more easily leave a sense of regret.''} 
This feedback prompted her to reflect on why the idea came up and finally clarify that evoking a sense of regret had been a core intention guiding her script development.

\revise{\subsubsection{Shifting creative focus by offloading routine workload to AI} By taking on routine and labor-intensive tasks, AI enabled screenwriters to shift their focus from technical execution to creative expression. Participants described how AI reliably handled standardized and information-heavy work, such as \textit{``preliminary investigation''} and \textit{``handling miscellaneous work''} (P9, P13), as well as formatting and structural preparation. This freed them to concentrate on original contributions. P8 explained that his role was to define \textit{``the creative framework and main narrative direction''} while AI \textit{``filled in some details within the framework.''} This redistribution also shifted attention away from efficiency and correctness toward thematic, narrative, and affective concerns. P1 planned to delegate summarization so he could \textit{``focus more on weaving the plot.''} P7 noted that outsourcing theoretical synthesis allowed him to connect theory with practical storytelling. P18 added that the credibility of characters' \textit{``emotions and motivations''} still required human judgment, while logical coherence could be largely supported by AI.}

\subsubsection{Iterating more flexibly with AI enabling rapid alternatives}  
Because the cost of exploring alternatives was greatly reduced, participants felt freer to abandon ideas and pursue new directions without being constrained by sunk costs. For instance, P7 noted that in the past, after investing significant time in one idea, she would \textit{``just execute the previous idea,''} even if it was unsatisfying. With AI, however, she felt encouraged to abandon earlier drafts and experiment with new trajectories. P12 similarly described AI as a high-efficiency \textit{``preview function''} that allowed her to quickly test stylistic changes and decide \textit{``whether to pursue this direction,''} thus supporting on-the-fly adjustments.

\subsubsection{Adding self-checking stages through AI-supported validation}  
Beyond reshaping existing phases, participants introduced new checkpoints into their workflows, using AI as a self-checking mechanism before seeking external feedback.  
In the idea stage, AI was used to check idea originality.  
P15 described deliberately challenging himself to \textit{``write something different from what AI would normally produce''} to ensure novelty.  
At the drafting stage, AI acted as a surrogate audience to anticipate market or audience responses.  
P14, for instance, asked AI to \textit{``assess whether this script could sell''} as a way to pre-evaluate commercial viability.  
Before sharing the draft with others, P14 also asked AI to
\textit{``run through multiple checks in advance, so that when I show it to others, it is already more polished.''}  
Together, these practices show how AI created new opportunities for verification, embedding self-checking as a recurring stage across different phases of the creative workflow.  

\subsection{Screenwriters' Concerns of Skill Sets}\label{sec:Ability}
Sec.~\ref{sec:Workflow} implies that participants chose to delegate tasks to AI for effortless yet high-quality screenplay creation.
Through reflections on task delegation, participants became aware of how it may lead to the erosion of certain capabilities and thus generate concerns.
It represents an active exercise of agency: screenwriters deepened their understanding of the consequences of their own decisions and became awareness of their impact on skill development. 

First, participants acknowledged that actively outsourcing demanding phases of ideation and evaluation to AI weakened their depth of thinking. Because AI produced rapid responses, many reported skipping the uncertain but essential stages of intuition, exploration, and deliberation that typically inform early creative reasoning. P2 admitted that he often bypassed foundational intuitive steps once AI provided an instant answer.
Participants further noted that such shortcutting had consequences beyond cognitive effort. P1 reflected that repeated exposure to AI's \textit{``dry''} structural outputs dulled his sensitivity to aesthetic qualities such as rhythm, mood, music, and mise-en-scène, which he considered crucial to his broader artistic judgment. At a higher level, concerns extended to independent reasoning: P15 worried that when AI-generated ideas were beyond his understanding, continued dependence would constrain his ability to judge validity and coherence, effectively shifting evaluative authority from humans to AI.

Participants also noted that delegating tasks to AI reduced their opportunities to practice the skills required for executing their ideas.
Automating drafting, revision, and information synthesis meant that the hands-on, iterative work of crafting dialogue, refining pacing, and polishing scenes occurred less frequently.
In this context, P6 and P9 described relying on AI for \textit{``filling in word count''} rather than investing their own effort in translating ideas into concrete written form.
P16 confessed that once AI could \textit{``simplify and extract key points,''} he became \textit{``so lazy that I didn't want to carefully read all those texts myself.''} 
As a result, substantial portions of the execution process were shifted from the writer to the system, limiting opportunities to practice.

%Participants also noted that delegating tasks to AI reduced their opportunities to practice skills required for executing their ideas. 
%Automating drafting, revision, and information synthesis meant that the hands-on, iterative work of crafting dialogues, refining pacing, and polishing scenes occurred less frequently. 
%P6 and P9 described relying on AI for \textit{``filling in word count''} rather than engaging in substantive thought. 
%P16 confessed that once AI could \textit{``simplify and extract key points,''} he became \textit{``so lazy that I didn't want to carefully read all those texts myself.''} 
% P6 and P9 admitted using AI as a shortcut for repetitive drafting, while
% These reflections highlight that losing exposure to the iterative mechanical work of writing also diminishes opportunities to maintain fluency, accuracy, and stylistic nuance.

\subsection{Screenwriters' Transformation of Cognition}\label{sec:Cognition}
Through sustained interaction with AI, participants gradually reorganized their cognitive models of creation. Rather than relying solely on intuition or habitual approaches, they began to internalize AI's structural thinking pathway and take diverse and objective viewpoints to examine their creation. 

Recently, LLMs with reasoning capability emerged, such as the DeepSeek Reasoner model.
Their ability to externalize the structural thinking process for screenwriting encourages human creators to move from amorphous intuition toward systematic analysis. 
P1 remarked that he learned to \textit{``look at stories with a kind of deconstructive thinking,''} and P2 reflected that through repeatedly judging AI's structural suggestions, he \textit{``became more proficient at mastering the structure.''} P9 believed that extended co-creation would result in \textit{``structural and functional thinking being deeply rooted''} in her mind. 

Furthermore, AI's diverse output encouraged creators to evaluate their work from external and multiple angles, rather than personal subjectivity during co-creation. P14 explained that he had \textit{``learned this kind of objective perspective''} from AI and integrated it into his workflow, and P17 highlighted gaining the ability to \textit{``look from both macro and micro angles''} when analyzing stories. Others noted that AI's presentation of alternatives fostered screenwriters' openness to different narrative approaches. P12 said, \textit{``It was like a pre-run: AI quickly showed alternative routes, which helped me decide whether to shift the story's path... Over time it broadened my scope and reminded me there are always other angles, and that the structure can be reconfigured.''} P11 expressed that, \textit{``Its associations weren't the same as mine, but I accepted them; they came from a different angle and felt more novel.''}

\subsection{Screenwriters' Reconsideration of AI Roles}\label{sec:role}
% \subsubsection{}  
Through sustained interaction, screenwriters gradually reframed AI from a technical tool to a partner in screenwriting.
First, screenwriters emphasized AI as a dependable working colleague that accompanied them throughout the process, patiently responding to every query and providing consistent support.  
P2 remarked that \textit{``basically, as soon as I start writing, I need AI,''} and P12 emphasized that AI acted as a \textit{``dialogue partner''} that \textit{``constantly provides me with information, no matter what I ask.''} Such comments portrayed AI as an always available partner to rely on. 

Second, beyond assistance, some screenwriters reframed AI as a credible creative competitor that stimulated self-examination and growth. 
When their ideas surpassed AI's, they felt more confident in ideas' originality.
When their thinking resembled AI's, screenwriters sometimes regarded it as ordinary and pushed themselves to think further. 
P17 described adopting this stance explicitly, noting that he would \textit{``first examine whether my own ideas are equally fresh''} before writing. Similarly, P18 emphasized that observing AI's thinking paths made her question whether her own associations had fallen into a fixed path similar to AI's, motivating her to break free from habitual patterns. 

Third, screenwriters also envisioned AI as an emotional support partner, providing encouragement and companionship in a solitary and often stressful practice. 
P5 described the most meaningful dimension as the \textit{``sense of companionship''} AI offered, highlighting how its steady encouragement and positive feedback helped him \textit{``keep going and finish writing.''} He further explained that creators are often fragile and anxious, and AI's supportive presence in practice alleviates this psychological pressure. Similarly, P19 likened AI to a \textit{``friend''} who wrote alongside her, noting that while she retained the lead, AI's supportive presence reduced feelings of isolation and provided reassurance. 

\section{RQ3 Findings: Calling for Support to Human Agency in Human-AI Co-Creation Systems}\label{sec:RQ3}

\revise{Grounded in recurring co-creation experiences with AI, as described in \textbf{RQ1} and \textbf{RQ2}, participants articulated expectations for future human–AI co-creation systems (\textbf{RQ3}).
Rather than viewing AI merely as a generative assistant, participants called for systems that engage with different properties of human agency throughout the creative process.
Building on this view, we identified four modes through which AI could support the agency: enabling clearer intention formulation and shared planning (plan-making partner), fostering foresight through proactive exploration (proactive and farsighted thinker), maintaining creative control through self-monitoring and feedback (responsible gatekeeper), and scaffolding long-term reflection and growth (reflective mentor).
Together, these modes help users sustain awareness, adaptability, and authorship within evolving human–AI collaborations.}

\begin{figure*}
%[H]
 \centering         
\includegraphics[width=1\textwidth]{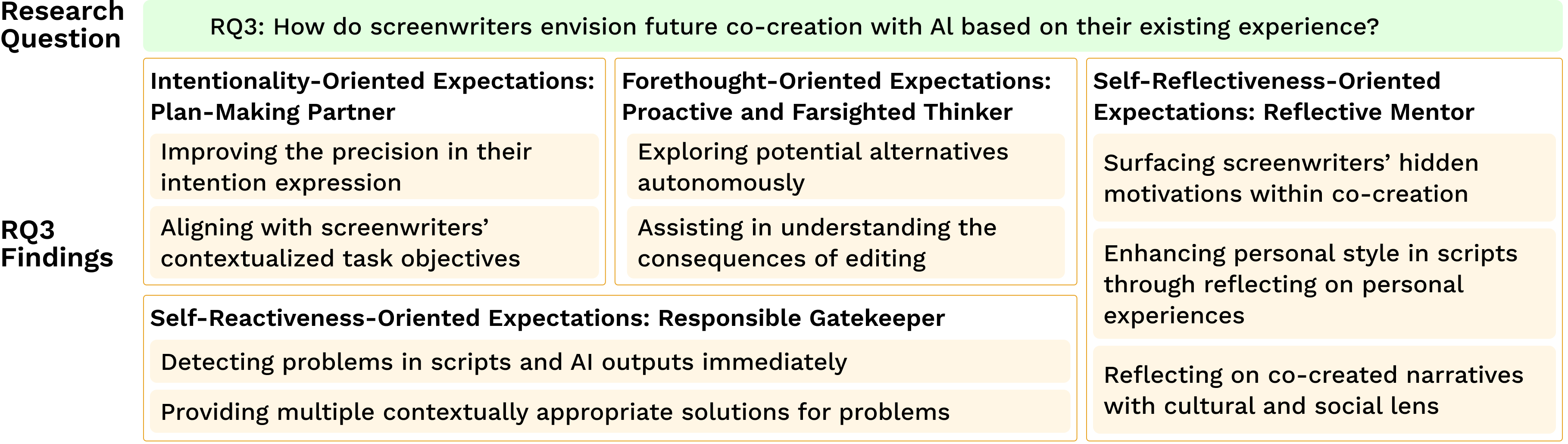} 
 \caption{Findings for RQ3. \rerevise{This figure maps participants' envisioned AI roles onto four human agency-oriented dimensions: intentionality, forethought, self-reactiveness, and self-reflectiveness. It illustrates screenwriters' expectations for how future AI systems may support different aspects of human agency in co-creation, including acting as plan-making partners to support intentionality, as proactive and farsighted thinkers to enhance forethought, as responsible gatekeepers for self-reactiveness, and as reflective mentors to facilitate self-reflection.}}
 \label{RQ3-fingdings}
\Description{}
 \end{figure*}

\subsection{Intentionality-Oriented Expectations: Plan-Making Partner}

To enhance planning co-creation tasks with their \textit{intentionality}, participants proposed that AI could better serve as a plan-making partner by supporting screenwriters in formulating clearer task objectives and establishing shared ground with the system from the outset. They highlighted two essential directions for such support.

First, participants expect that AI could help them improve the precision in their intention expression.
They recognized that vague or inconsistent descriptions of their plans to AI
often undermined AI's execution results, leaving them to rely on reactive adjustments later. As several admitted, \textit{``maybe I didn't express it clearly''} (P7) or \textit{``my prompt was not specific enough''} (P2).
P18 wished AI could \textit{``automatically detect logical problems in my input prompt''} and even \textit{``reject flawed loopholes''} before generation (related to Sec.~\ref{sec:Logical}). 
They hoped AI could serve as a pre-emptive check, helping them refine and clarify their instructions before execution. By reducing ambiguity at the outset, such support would not only improve alignment in task objectives but also lessen the compensatory burden on \textit{self-reactiveness} and create clearer anchors for later reflection.  

Second, AI should be able to align better with screenwriters' contextualized task objectives in the planning stage through better context communication.
% Another reason for unsatisfactory execution by AI is that it cannot sufficiently understand user intent and plan without context information.
% To address the issue, our participants expected a better alignment between humans and AI in the planning stage through better context communication.
For example, P1 wished AI could reference his \textit{``personal experiences and memories''} to generate more relevant suggestions, aiming for narrative alignment as his evaluative focus (related to the Sec \ref{sec:Alignment}), and P3 emphasized drawing on \textit{``real-life cases or news''} to ensure factual accuracy (related to the Sec. \ref{sec:Accuracy}). 
Grounding detailed intentions in their context during the intention-building process could reduce the need to revise AI outputs, thereby enabling more efficient and time-saving human-AI co-creation.

\subsection{Forethought-Oriented Expectations: Proactive and Farsighted Thinker}

To facilitate \textit{forethought}, AI is expected to act as a proactive and farsighted thinker that advances creative foresight through proactive exploration.  
In this way, they can help screenwriters better discover and even predict potential paths and branches in narratives.

First, screenwriters would like AI to explore more potential alternatives autonomously to expand the horizon of creative possibilities. 
They criticized current systems for mechanically \textit{``stitching''} elements together, with P18 wishing for more thoughtful and imaginative fusion rather than what he described as AI's tendency to \textit{``brutally stitch two elements together.''} P4 further called for algorithms that could \textit{``actively and spontaneously innovate on content''} so outputs would not \textit{``feel plagiarized.''} 
To meet this expectation, participants envisioned AI capable of proactive exploration, suggesting alternative pathways or scenarios before writers invested effort in drafting. P17, for instance, wanted AI to recommend \textit{``which scenario would best present my story,''} thus providing creative options beyond those he would have considered independently.

They also hoped that AI could \textit{help them anticipate the consequences of any modification}.
P12 imagined AI anticipating \textit{``the whole game''} by predicting the ripple effects of a single narrative decision across the broader story system.
P11 and P14 proposed a blueprint-style visualization that allows screenwriters to trace how small narrative choices influence global trajectories and branches.  
It aligns with creative objectives in \textit{forethought} (Sec.~\ref{sec:forethought}), asking AI to act as a catalyst for forward-looking creation beyond an executor.

\subsection{Self-Reactiveness-Oriented Expectations: Responsible Gatekeeper}

To facilitate \textit{self-reactiveness}, participants would like AI to serve as a responsible gatekeeper that could share the workload of monitoring and proposing methods to enhance outcomes. 

Specifically, AI should be able to immediately detect problems in scripts, such as logical inconsistencies, pacing problems, or stylistic mismatches, before such flaws accumulated. 
Participants emphasized that this should include not only post-hoc feedback but also proactive self-checking mechanisms, where the AI inspects its own outputs before presenting them to users. 
P9 mentioned \textit{``AI should check itself first before providing the output for users.''} Likewise, P7 appreciated the potential of such immediate regulation: \textit{``If it can catch the weak points the moment they appear and adjust automatically, I don't need to waste energy fixing them later.''} This real-time scrutiny would save writers from extensive downstream revisions and help maintain narrative coherence. For instance, P5 wished AI could \textit{``flag the awkward jumps in tone before I move too far forward,''} underscoring the importance of timely feedback in co-creation.
Beyond merely flagging problems, participants expected AI to provide multiple contextually appropriate solutions. 
P1 expected AI to recognize her current creative context as an artistic adaptation from a homosexual perspective and to propose multiple alternative directions that align with this theme in a coherent manner. As she explained, \textit{``If the AI finds that telling the story of \textit{Swan Lake} from a homosexual perspective seems far-fetched, then it should offer other stories, like A, B, C, or D, that could be interpreted from this angle.}'' 
% This expectation illustrates that screenwriters sought diverse alternatives that were contextually aligned with their ongoing creative vision.

\subsection{Self-Reflectiveness-Oriented Expectations: Reflective Mentor}\label{sec:Mentor}

\revise{To cultivate \textit{self-reflectiveness}, participants imagined AI as a reflective mentor that could scaffold long-term processes of reflection and reinterpretation. This expectation extended beyond immediate problem solving through \textit{self-reactiveness} toward helping them develop motivation and identity in co-creation. This view is consistent with prior work showing that reflection is central to human–AI creative collaboration~\cite{shen2025ideationweb, ford2022speculating, glinka2023critical}.}

\revise{First, participants hoped AI could help surface hidden motivations and creative drives in their collaborative process. They emphasized that reactive drafting support was insufficient and that AI should participate in uncovering latent considerations that guide decision-making. Building on the benefits of intent clarification described in Sec.~\ref{sec:Clarifying_intention}, they expected AI to reveal deeper impulses behind their choices. For example, P15 wished AI could identify \textit{``the user's original motivation when integrating two elements,''} and P2 expected it to \textit{``clarify my feelings... to help me uncover what exactly touches me in the ideas I generate.''}
Such assistance transforms screenwriters' instincts into awareness, helping them recognize deeper drives and strengthen their creative work.}

\revise{Second, participants also expected AI to prompt the recalling and reinterpreting of personal memory during co-creation. They wanted AI to guide them in revisiting lived experiences and overlooked details, turning memory into a generative resource rather than a passive backdrop. P17 wished AI could \textit{``guide me to better recall''} forgotten fragments and encourage exploration of \textit{``what if''} variations. This expectation extends from current practices described in Sec.~\ref{sec:adaption}, where even minor impressions could inspire new directions. By activating memory in this way, participants hoped to enrich narratives with more nuanced emotions, personal histories, and distinctive styles, ensuring that collaborative outputs retain the creator's individuality instead of defaulting to generic patterns.}

\revise{Finally, participants expected AI to contextualize co-created narratives within cultural and social frames in ways that foster reflection on their own positions and voices. Although AI already supplied factual or cultural references (Sec.~\ref{sec:Accuracy}), they hoped it would illuminate how their work might align with or diverge from broader social meanings. P7 expressed a desire for \textit{``details of genuine emotions that imagination alone cannot reach,''} and P3 suggested grounding insights not only in canonical works but also in \textit{``real cases in everyday life, like interviews with people and personal accounts.''} By situating narratives within wider cultural contexts, AI would encourage creators to reflect on their values and identities, thereby strengthening \textit{self-reflectiveness} more broadly at the intersection of narrative, lived experience, and society.}

\section{Discussion}
% In this section, we synthesize our findings into implications for creative practice (Sec.~\ref{sec:suggestion_screenwriter}), human–AI co-creation tool design (Sec.~\ref{sec:suggestion_tool}), and future research (Sec.~\ref{sec:suggestion_research}). 
This section begins with suggestions for screenwriting practice, tool design, and future research in human-AI co-creation (Sec.~\ref{sec:suggestions}).
We also present overarching takeaways for readers (Sec.~\ref{sec:Lesson}), discuss the potential risks (Sec.~\ref{sec:Risks}), and reflect on limitations with directions for future work (Sec.~\ref{sec:limitation}).

\subsection{Suggestions} \label{sec:suggestions}
In this section, we synthesize our findings into implications for creative practice (Sec.~\ref{sec:suggestion_screenwriter}), human–AI co-creation tool design (Sec.~\ref{sec:suggestion_tool}), and future research (Sec.~\ref{sec:suggestion_research}). 

\begin{figure*}
%[H]
 \centering         
\includegraphics[width=1\textwidth]{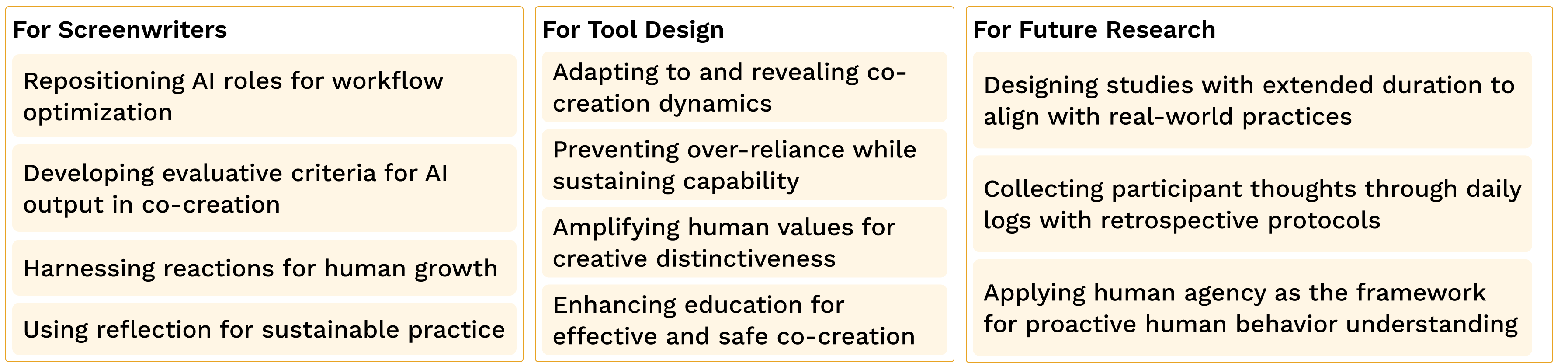} 
 \caption{Suggestions for Screenwriters, Tool Designers, and Future Researchers. Based on our findings, this figure summarizes a series of suggestions for various stakeholders in human-AI co-creation to understand and foster human development.}
 \label{suggestion}
\Description{}
 \end{figure*}
 
\subsubsection{For Screenwriters}\label{sec:suggestion_screenwriter} 
\rerevise{Grounded in our findings, we outline two types of suggestions for screenwriters. 
The first type requires screenwriters to be aware of these ongoing changes and navigate them from now on, including repositioning AI roles for workflow optimization and developing evaluative criteria for AI output in co-creation.
The other type serves as early reminders for screenwriters to consider their future practices to amplify their advantages and mitigate potential risks~\cite{10.1145/3613904.3642529}, including harnessing reactions for human growth and using reflection for sustainable practice.}

\textbf{Repositioning AI roles for workflow optimization.}
Our research identified the emerging tasks for AI application (Sec.~\ref{sec:intentionality}) and the trend of treating AI as partners rather than tools (Sec.~\ref{sec:role}).
The repositioning of AI roles will continue as we are witnessing a fast development of AI technologies~\cite{10.1145/3544548.3580782}.
To catch up with the trend, screenwriters need to actively reflect on and develop how they can combine the new positive and negative aspects of AI with their practices, through actively iterating the plans for using AI and rotating AI between roles across tasks.
In this way, screenwriters can continuously update their workflow for effectiveness and effortless co-creation with AI while ensuring their values.

\textbf{Developing evaluative criteria for AI output in co-creation.} 
Our findings show that screenwriters actively developed evaluative criteria to guide how they assessed and anticipated AI output during co-creation (Sec.~\ref{sec:forethought}). These criteria informed decisions made before invoking AI, such as whether and how to use AI for a goal, and also shaped writers’ immediate reactions to the generated output. 
In the future, writers should cultivate multi-level standards: pragmatic criteria for immediate usefulness and aesthetic criteria for preserving the unique human voice \cite{10.1145/3613904.3642625}. These standards should be periodically audited and updated to align with the development of AI models and fulfill the ultimate goal of safeguarding humans' unique intuition, taste, and judgment.

%As our findings reveal, screenwriters came up with various evaluative criteria to guide their foresight before applying AI (Sec.~\ref{sec:forethought}).
%These criteria also facilitated plan-making and immediate reaction to AI output.
% constructing evaluative criteria is crucial to provide perspectives for foresight (Sec.~\ref{sec:forethought}).
% Examining expected AI output with these criteria facilitates plan-making and immediate reaction to AI output.
%In the future, writers should cultivate multi-level standards: pragmatic criteria for immediate usefulness and aesthetic criteria for preserving unique human voice \cite{10.1145/3613904.3642625}. 
%These standards should be periodically audited and updated to align with the development of AI models and fulfill the ultimate goal of safeguarding humans' unique intuition, taste, and judgment.
% The key goal of these standards should remain as safeguarding humans' unique intuition, taste, and judgment. 
% To achieve the key goal, it is important to make detailed standards explicit for periodically auditing and updating.
% Making these explicit and periodically auditing decision-making helps maintain mastery while benefiting from AI's efficiency.

\textbf{Harnessing reactions for human growth.} 
Each response to AI output, whether adoption, adjustment, or avoidance (Sec.~\ref{sec:self-reactiveness}), should be treated as an opportunity to refine screenwriters' skills, including instructing AI actions, evaluating and criticizing creative products, and drafting scripts manually. 
Even AI failures can be turned into cornerstones for success by asking why they failed and what alternatives they reveal. 
Through continuously and actively responses, screenwriters can potentially mitigate the concern for skill erosion (Sec.~\ref{sec:Ability}) and gradually develop themselves.

\textbf{Using reflection for sustainable practice.} Reflection should include not only reviewing past human-AI collaborations but also proactive planning. Screenwriters should consider how AI has reshaped their strategies, confidence, workflow, abilities, and cognition (Sec.~\ref{sec:RQ2}), then project forward: What strengths to preserve? What vulnerabilities to guard against? What strategies ensure AI reinforces rather than undermines creative identity? Such reflection transforms temporary lessons into durable practices. The expectations articulated in Sec.~\ref{sec:RQ3} exemplify this rethinking, and future screenwriters should embrace it as a continuous discipline to sustain human distinctiveness in co-creation.

%\subsubsection{For Tool Design}\label{sec:suggestion_tool}
%\rerevise{In Sec.~\ref{sec:RQ3}, screenwriters articulate a wide range of practical needs for AI systems based on their current co-creation practices. Building on these practice-grounded findings, we propose higher-level design suggestions for future tools that facilitate broader human–AI co-evolution. While our analysis centers on screenwriting, these suggestions are grounded in creators' existing practices and extend beyond this context. In the following sections, we use the term \textit{creator} to refer not only to screenwriters but also to practitioners in other creative domains who engage in sustained co-creation with AI.}

\subsubsection{For Tool Design}\label{sec:suggestion_tool}
\rerevise{In Sec.~\ref{sec:RQ3}, we identified a range of practical needs articulated by screenwriters through their current co-creation practices with AI. Building on these findings, this section synthesizes higher-level design suggestions for future AI tools. 
We believe that our suggestions have the potential to benefit tools for \textit{creators} in both screenwriting and other creative domains.
}
% While our analysis centers on screenwriting, these suggestions extend beyond this context. Accordingly, we use the term \textit{creator} to refer to both screenwriters and practitioners in other creative domains who engage in sustained co-creation with AI.}

%While our analysis is situated in screenwriting, the suggestions are not limited to this context. Accordingly, in the following sections, we use the term \textit{creator} to refer to screenwriters as well as practitioners in other creative domains who engage in sustained co-creation with AI.}

\textbf{Adapting to and revealing co-creation dynamics.}  
Since co-creation is dynamic, tools must adapt as workflows, strategies, and expectations change. Designs should support users' flexibility and awareness of how human–AI relationships shift over time to support their self-reflectiveness. For example, systems could visualize long-term collaboration patterns, showing how reliance on AI fluctuates or how strategies (e.g., using AI for ideation vs. editing) reshape outcomes. Such longitudinal insights enable creators to track evolving practices and recalibrate collaboration strategies. Empirical work shows varying levels of AI assistance can have U-shaped effects on quality and productivity, highlighting the need to personalize scaffolding to user needs \cite{10.1145/3613904.3642134}. By revealing and responding to co-creation dynamics, tools can help users actively manage collaboration.
\revise{For example, motion graphics artists could track how their reliance on AI-generated transitions fluctuates across a project, helping them notice when manual animation practices diminish and adjust to sustain essential timing skills.}

\textbf{Preventing over-reliance while sustaining capability.}  
Over-reliance risks diminishing engagement and eroding evaluative judgment or narrative craft \cite{10.1145/3613904.3642625}. Beyond ad hoc distancing, tools should embed mechanisms that encouraging humans to exercise their agency instead of over-reliance. 
For instance, systems might ask humans to criticize and improve their current instructions for AI (\textit{self-reactiveness}) and archive critiques as a personalized critical thinking profile for long-term self-reflection.
Also, systems can stage intervals where users must make key decisions or problems to avoid skill erosion.
Such strategies transform distancing and practice into structured capability cultivation.  
\revise{
% They can also be applied to other creativity domains.
For example, a system could ask designers to revise and justify their own design briefs before the AI generates concepts, and later require them to choose between several conflicting design directions using their own criteria. These steps preserve intentional judgment and reasoning and still allow AI to support later refinement.}

\textbf{Amplifying human values for creative distinctiveness.}  
Besides, tools should sustain human distinctiveness rather than flattening expression into generic patterns. The goal is not only preserving style but also foregrounding values and worldviews that shape creative work. Systems could let creators define evolving value maps, including motifs, ethical stances, and emotional tones, and visualize how these shift across drafts.
Reflective dashboards might compare the creators' value profile with broader community trends, showing what is distinctive and what risks assimilation into AI-driven conventions. In this way, tools function as both productivity aids and amplifiers of human distinctiveness. 
\revise{
% These dashboards also have potentials to benefite other creative fields. 
For example, painters can track how their use of cultural symbols, color philosophies, or recurring themes evolves across iterations with these dashboards. They can compare these elements with prevailing AI-generated styles to reveal where the artist's work remains distinctive and where it begins to drift toward homogenized AI aesthetics, helping them stay grounded in their own worldview.}

\revise{\textbf{Enhancing education for effective and safe co-creation.}
Our research reveals developments in workflow, confidence, and delegation accumulate gradually (Secs.~\ref{sec:Workflow}–\ref{sec:Cognition}) at an individual level.
Tool designers can also take the responsibility as mentors to train or educate creators to ensure that they can develop in the right direction and even accelerate beneficial development.
Designers can host tutorials or seminars for the tool usage with information about how the tool can facilitate their development through exercising agency.
For example, for screenwriters, they can share how to should critically assess and react to AI ideas for independent thinking in the brainstorming stage.
% can share potential usage in the brainstorming stage, such as using AI to diversify ideas while preserving personal identities.
When hosting such sessions for creators in other domains, such as illustrators or concept designers, workshops could demonstrate how to set feasible plans for AI to effectively boost artifact creation.
% how to effectively boost artifact creation with AI assistance under humans' directions.
%\haotian{Yuying, please add another example for a different domain.}
They should also introduce how to handle risks in human-AI co-creation, such as skill erosion, reduced personal voice, and uncritical acceptance, through these occasions, and also add warning messages and mechanisms in tools for educational purposes.}

\subsubsection{For Future Research}\label{sec:suggestion_research}
Creative processes are inherently longitudinal and iterative, involving many mental activities not directly observable from surface behaviors. Understanding practical human–AI co-creation, therefore, requires research designs that capture hidden, evolving, and subjective dynamics over time. \rerevise{Grounded in our research experience, we highlight three suggestions}.

\textbf{Designing studies with extended duration to align with real-world practices.}  
Screenwriting is often a complex project involving multiple stages, such as ideation, outline drafting, scene development, and dialogue creation~\cite{tang2025understanding}.
It is hard to be 
meaningfully studied in short laboratory sessions, as confirmed by our pilot observations.
We found that creators would rely more on AI for efficiency when given limited time, diminishing their own personal values and making the study biased.
As a result, we employed a two-week period to allow screenwriters to sufficiently engage in complete cycles of human-AI co-creation, aligning the study more closely with real-world practice. 
Other co-creation domains may also require more time beyond hour-long lab study sessions, such as music composition and video production. 
Future research should carefully consider task duration and allocate sufficient time to capture the temporal unfolding of creativity, ensuring ecological validity and reliable results.

\textbf{Collecting participant thoughts through daily logs with retrospective protocols.}  
% Concurrent think-aloud risked disrupting writing flow, so participants instead kept daily logs, followed by retrospective think-aloud and interviews. 
When designing our study, one critical issue was how to understand the inner thoughts of screenwriters when exercising agency, such as how they decided tasks for AI and evaluated AI output with their criteria.
The issue was severe in a two-week-long study, as participants might write the screenplay at any time they preferred, making real-time think-aloud hard to capture by researchers.
Also, we worried that real-time think-aloud might interfere with their creation flow.
On the other hand, fully relying on retrospective reporting by participants might risk the loss of details, such as intermediate ideas and struggles.
To address the issue, we adopted a hybrid approach: participants kept concise daily logs in spreadsheets; at the end of the study, retrospective think-alouds drawing on the logs, followed by semi-structured interviews, elicited holistic reflections. This design preserved the authenticity of the creative process and enabled richer accounts of thought trajectories and decision-making~\cite{10.1145/3706599.3706676}.
Our design can serve as references for future agency-related study design.

\textbf{Applying human agency as the framework for proactive human behavior understanding.}  
We analyzed co-creation through the lens of human agency, focusing on \textit{intentionality}, \textit{forethought}, \textit{self-reactiveness}, and \textit{self-reflectiveness}. Within human–AI interaction (HAI), agency offers a critical perspective for explaining how humans not only respond to AI outputs but also actively regulate, anticipate, and reflect on their collaboration with AI. Prior research has often conceptualized agency from a static standpoint, emphasizing passive influence or limiting analysis to dimensions of control and decision support~\cite{10.1145/3715928.3737481}. By contrast, our approach foregrounds agency as a dynamic and evolving capacity, highlighting how creators continually shape the trajectory of co-creation. This perspective provides more nuanced empirical insights into the interplay between human initiative and AI assistance, and underscores the importance of designing systems that recognize and scaffold proactive human behavior.

\subsection{Takeaways}\label{sec:Lesson}

Generative AI has demonstrated substantial power in generating texts, such as poems~\cite{10.1145/3610591.3616427, 10.1145/3610537.3622957} and stories~\cite{chung2022talebrush}, and visuals~\cite{bansal2024revolutionizing}, such as images~\cite{10.1145/3678884.3681890} and videos~\cite{zhou2024survey}.
To embrace and shape the potential game changer, it is crucial for human creators to actively develop and adapt to the new paradigm of human-AI co-creation.
At the time point, when the first batch of generative AI models had been released for three years, we explored and provided a snapshot of how creative workers, particularly screenwriters, have grown to master the usage of AI with their agency: how they set intentions, anticipated outcomes, regulated actions, and redefined their collaboration with AI based on reflections on their practices.
Throughout our research, we identified a series of insights. 
  
First, human-AI co-creation is not characterized by passive consumption of machine outputs but by the active exercise of human agency. Screenwriters deliberately set intentions, applied evaluative foresight, and regulated interactions in real time, showing that humans remain central in steering AI for their outcomes. 

Second, screenwriters have evolved to actively embrace co-creation with AI with their adaptation and reflection. They developed new paradigms to amplify human value with AI participation through reconfiguring their capabilities, updating their workflows, and enhancing their cognition.

Third, participants envisioned AI as a co-evolving partner that supports goal-setting, acts predictably, collaborates in evaluation rather than generation, and mentors long-term growth. These expectations highlight that AI should not only solve problems but also nurture human agency and development.  

Combining the three findings, our study also underscores a paradigm shift: co-creation is not only about producing content more efficiently but also about \rerevise{developing their capacities}, expressing values, and transforming identity through sustained human-AI collaboration. 
It highlights that the ultimate target for research on human-AI co-creation should be more diverse beyond efficiency and product quality~\cite{rhys2025beyond}.
For example, co-creation tools should cater for the growth and development of human themselves, such as increasing their self-efficacy and enhancing the expression of their inner thoughts through creative practices.

Beyond revealing how humans engage with AI in creative work, our findings show that users' needs for AI evolve dynamically, shifting from seeking functional assistance and efficient outputs to valuing emotional and reflective support through sustained interaction (see Sec.~\ref{sec:Confidence} \& Sec.~\ref{sec:role}). For example, screenwriters increasingly viewed AI not only as a means to improve efficiency but also as a collaborator that broadened their creative capacity (refer to Sec.~\ref{sec:Confidence}). Additionally, many screenwriters further appreciated AI's steady encouragement and companionship, describing it as a presence that helped them cope with pressure and maintain emotional resilience during solitary writing (refer to Sec.~\ref{sec:role}). Together, these developments indicate a gradual reconfiguration of human–AI relationships, where technical assistance becomes closely connected with emotional support and reflective engagement. 

Looking forward, these insights collectively suggest that we should obtain both in-depth understanding about the impact of AI on humans and humans' needs for AI concurrently in parallel to catch up with the evolution of both humans and AI and set a meaningful agenda for research and development in the next phase.
The framework of human agency offers a promising foundation for this work, as it helps characterize how people formulate intentions, anticipate outcomes, regulate actions, and reflect on their evolving practice in pursuit of their goals. 

\rerevise{Guided by the framework, our study deepens the understanding of mechanisms and factors to shape humans' AI usage and thinking modes, which can hardly be captured by studies that targets at snapshots of human-AI co-creation.
For example, our study reveals the diverse and mixed strategies that drove the formation of working modes in co-creation (Sec.~\ref{sec:RQ1} \& Sec.~\ref{sec:adaption}) and also human development led by continuous interactions with AI, such as the increasing confidence led by AI's external verification and emotional support (Sec.~\ref{sec:Confidence}) and the internalization of AI's thinking modes (Sec.~\ref{sec:Cognition}).
Building upon our findings, future research can generate a deeper understanding of human changes and needs.
Then, researchers can propose richer strategies for supporting humans as they advance in long-term symbiosis with AI, from creative professions to everyday contexts, and from individual growth to societal change.}

\subsection{\revise{Risks Underlying Human Development in Human–AI Co-creation}}\label{sec:Risks}
\revise{Across the study, participants described various forms of development, yet none unfolded in a purely positive or purely negative direction. Instead, each form of development may contain both positive impacts and negative ones, or even risks. Though our participants are mostly optimistic about these changes, we would like to extend our discussion on potential risks to warn future creators and researchers.}
\revise{For instance, participants reported an increase in confidence when AI offered steady knowledge or validating viewpoints (Sec.~\ref{sec:Confidence}). This growth encouraged bolder choices and reduced hesitation. However, heightened confidence sometimes may narrow the space for self-checking, increasing the chance of accepting AI suggestions too quickly and producing moments of confusion about whether a decision reflected their own judgment or AI's influence. 
Workflow reconfigurations enabled faster iteration and clearer direction-setting (Sec.~\ref{sec:Ability}), but they may also reduce early intuition-driven exploration by involving AI for ideation. It raises concerns about long-term skill atrophy when foundational practices become less exercised. Delegating demanding tasks helped free cognitive resources for higher-level reasoning (Sec.~\ref{sec:Workflow}). 
It is also possible to increase reliance on AI and created early signs of dependency that might accumulate across sustained use. These examples indicate that development in human–AI co-creation should be understood as a dynamic balance between benefits and costs.}

\revise{As the field moves forward, it is important to both enhance our in-depth understanding of negative impacts or risks led by the observed changes mentioned in Sec.~\ref{sec:RQ2} and further consider how to help creators amplify positive impacts and mitigate the negative ones. 
To gain more insights about negative impacts, it can be helpful to explore methods beyond interviews.
For example, to examine deskilling, it is possible to set up experiments to periodically ask creators to work without AI and trace their performances in the long run.
Also, to reveal potential emotional dependency levels, it might be essential to develop corresponding scales and questionnaires. From the tool design perspective, one important aspect is to both enhance efficiency and long-term creative capacity. 
Although AI can accelerate ideation and structuring, this acceleration may reduce the time spent reasoning through difficult choices or practicing craft, making skill atrophy more likely to arise gradually (Sec.~\ref{sec:Workflow} \& Sec.~\ref{sec:Ability}). Future systems need to account for how rapid support may unintentionally shorten stages that nurture intuition, hands-on experimentation, and rigorous self-evaluation. 
% Attending to this balance is essential for sustaining creative capacity across long-term collaboration. 
Another consideration relates to creative expansion and the emotional dynamics of collaboration. AI's ability to broaden possibilities or offer emotional reassurance can encourage exploration (Sec.~\ref {sec:Cognition} \& Sec.~\ref{sec:role}) , yet it may also shift narratives in unintended directions or foster dependency when writers begin to rely on AI's presence for reassurance or stability. Designing for this space requires sensitivity to how emotional support interacts with authorship, and how expanded creative options can introduce fatigue when writers must repeatedly disentangle their own voice from AI's influence. Recognizing these complexities will help guide future work toward systems that support creators without overshadowing or overriding their long-term development.}

\subsection{Limitations and Future Work}\label{sec:limitation}
This study advances understanding of how screenwriters leverage human agency to engage proactively in co-creation with AI, but several limitations and directions for future work remain.

\subsubsection{Limitations.}
First, we asked participants to complete a five-minute short film screenplay within two weeks, which enabled us to capture a full creative workflow and approximate authentic short-form screenwriting practices. Other formats, such as feature-length films or television series, may involve different uses of AI and present distinct challenges, but exploring these would require larger-scale and longer-term studies to capture the extended dynamics of co-creation. Second, the screenplays developed in our study were not staged or produced. Although this choice allowed us to focus on the creative process itself, it also meant that practical considerations such as performability, production constraints, and collaboration with other stakeholders were not addressed. Future work could incorporate these real-world factors to enrich the understanding of how agency is exercised in actual production contexts. \revise{Third, although snowball sampling facilitated access to professional screenwriters, it introduces potential selection bias and may limit the representativeness of participants' perspectives~\cite{goodman1961snowball}. And our sample shared relatively similar professional and cultural backgrounds, which may have shaped their perceptions of AI collaboration.
To address the problem, future research could recruit a diverse range of practitioners through other recruitment methods, such as reaching out to a more diverse group of screenwriters through emails or setting up a worldwide research network for participant recruitment.
Having participants across different regions, creative industries, and experience levels can further support a comparative study of agency across multiple dimensions with behavior analysis.} 
\revise{Finally, our study focuses on screenwriters who engaged in co-creation with AI. Recruiting participants with prior experience aligns with our goal of examining development, as meaningful development presumes some familiarity with co-creation. Within this group, we observed screenwriters who showed clear forms of development through continued interaction as well as those who showed little development, as discussed in Sec.~\ref{sec:disconnected}.}

\subsubsection{Future Work.}
\revise{Looking ahead, several directions emerge from the scope of this study. 
First, our analysis centered on screenwriters who engaged directly in co-creation with AI, and we used human agency as an analytical lens to describe how users developed through sustained practice. Future research could extend this work by operationalizing agency as a measurable construct and building a mechanistic account of how humans evolve over time. Such a model could explain transitions in agency, outline possible stages or mechanisms, and examine how factors such as user activeness, prior experience, interaction depth, genre, and task context shape this development.
Measuring agency level can also help us distinguish creators who actively exercise intentionality, forethought, self-reactiveness, and self-reflectiveness and those who are less active.
By contrasting their behaviors and continuous development, it is possible for us to gain more insights about how to facilitate human-AI co-creation for different creators in a long run.
}
\revise{Second, future work could compare different depths or forms of interaction to understand how varying configurations of collaboration influence the development of agency and affect creative decision making~\cite{10.1145/3613905.3651042, 10.1145/3706598.3713626}.} 
\revise{In addition, as our findings indicate that sustained interaction can lead to emotional connection or companionship dynamics, future studies could investigate the behavioral and computational processes that support these shifts and examine how AI systems adjust to users' evolving emotional needs.}
Finally, future research could also extend to creative domains such as game design, advertising, and education, which may surface different ways in which creators exercise and cultivate agency in co-creation. System design should also consider how to support sustained human agency by scaffolding intentionality, forethought, self-reactiveness, and self-reflectiveness across diverse professional contexts. Taken together, these directions can contribute to a broader and more sustainable understanding of long-term human–AI collaboration.

\section{Conclusion}
To understand how screenwriters exercise their agency to develop their co-creation with AI, we conducted a two-week study with 19 professional screenwriters. The study combined co-creation sessions, retrospective think-aloud protocols, and semi-structured interviews. Using Bandura's human agency, our analysis showed that screenwriters deliberately set plans, evaluated AI outputs against professional standards, and responded through immediate judgments and regulatory actions. Through reflection, they developed evolving strategies such as transforming disruptions into creative leverage and identifying disconnected effects without evolving impact. They also enhanced their confidence and expressed concerns about the erosion of their skill sets. In addition, they restructured workflows to integrate AI more flexibly and cultivated new cognitive practices that deepened multi-perspective thinking. Finally, screenwriters articulated ideal AI systems that better support the four properties of human agency. Based on these findings, we provided suggestions for screenwriters on exercising human agency in human-AI co-creation, for future tool design, and for research on professional creative practices. \revise{We also discuss the underlying risk of these human developments. In the future, we hope to continue investigating humans' usage of AI through their agency with long-term observation studies in diverse creative domains.}

\begin{acks}
This work was partially supported by the Research Grants Council of the Hong Kong Special Administrative Region under the General Research Fund (GRF) (Grant No. 16218724 and No. 16207923) and is part of the AFMR collaboration supported by Microsoft Research. We extend our gratitude to Jina Suh, Gonzalo Ramos, Jiaxiong Hu, and Yu-Zhe Shi for their valuable support and discussions on this work. We also sincerely thank all participants for their time and contributions. Finally, we greatly appreciate the reviewers for their insightful and constructive feedback.
\end{acks}

\bibliographystyle{ACM-Reference-Format}  % 或其他样式
\bibliography{main}  % references.bib 是你的 bib 文件名

\end{document}